\documentstyle[seceq,epsf,preprint]{ptptex}

\def\vecphi{\mbox{\boldmath $\varphi$}}
\def\veca{\mbox{\boldmath $a$}}

\title{
Superconvergence of period doubling cascade
in trapezoid maps 
}
\subtitle{Its Rigorous proof
 and superconvergence
of period doubling cascade starting from
a period $p$ solution}    

\author{
Tatsuya {\sc Uezu}
}

\inst{
Graduate School of Human Culture, \\
Nara Women's University, Nara 630-8506}

\recdate{
}

\abst{
In the symmetric and the asymmetric trapezoid maps,
as a slope 
$a$ of the trapezoid 
is increased, the period doubling cascade occurs and 
 the symbolic sequence of periodic points is 
the Metropolis-Stein-Stein sequence $R^{*m}$ and
the convergence of the onset point $a_m$ of the
period $2^m$ solution to the 
accumulation point $a_c$ is exponentially fast.
In the previous paper, we proved these results.\par
In this paper, 
we give the detailed description of 
the proof on the results.
Rigorously, we show that 
\begin{eqnarray*}
  \epsilon _m &=&
\frac{ b a_c ^{-2^m} \gamma ^{- \zeta _m }}
{G'_{\infty} (a_c)}(1 + h_m),\\
\delta_m & =& \gamma ^{(-1)^m/3 }
(a_c \gamma ^{2/3}) ^{2^m} (1+l_m),\\
\lim_{m \rightarrow \infty} h_m&=&0, \\
\lim_{m \rightarrow \infty} l_m & =& 0,
\end{eqnarray*}
where $\epsilon _m \equiv a_c - a_m, 
\delta_m \equiv \frac{\epsilon_m-\epsilon_{m+1}}
{\epsilon_{m+1}-\epsilon_{m+2}}$,  $b$ and $\gamma$ 
are the smaller size of the trapezoid 
and the ratio of its two slopes, respectively.
$\gamma = 1$ corresponds to the symmetric trapezoid.
Further, we study the period doubling cascade
starting from period $p(\ge 3)$ solution.
We show 
\begin{eqnarray*}
\epsilon _m & \propto & 
(\gamma ^2 a_{p,c}^p) ^{-2^m}
 \gamma ^{ \zeta _m },\\
 \delta_m &\simeq & 
\gamma ^{(-1)^{m-1}/3 }
 (a_{p,c}^p \gamma ^{4/3}) ^{2^m},
\end{eqnarray*}
where $a_{p,c}$ is the accumulation point of the onset of
the period $p \times 2^m$ solution.

}

\begin{document}

\maketitle

\section{Introduction}
It is well known that in a class of one-dimensional map,
as a parameter is changed, the period doubling 
bifurcation cascades,
 and there exist several universal properties
\cite{Feigen 78}.
One of the universal quantities is the so called Feigenbaum
constant $\delta$
and it depends only on the exponent $z$ characterizing the
behavior of the map around the critical point.

 About a decade ago, there had been controversy on the
$z \rightarrow \infty$ limit of $\delta(z)$.
As a typical example, the map 
$ x_{n+1}=f(x_n)= 1- a |x_n |^z $
has been extensively studied and
two different values were conjectured for this limit,
one is  finite and the other is infinity
\cite{Weele 86,Hauser 84}.
This problem was solved by 
J. P. Eckmann and H. Epstein and
 they proved that the limit is finite
\cite{Eckmann 90}.

In the previous paper\cite{Ue 98}, 
instead of considering a map 
with finite $z$ and
taking $z \rightarrow \infty$, 
we investigated a map with $z = \infty$.
As such  maps, we treated the symmetric and 
 asymmetric trapezoid maps.

\begin{figure}
     \centerline{\epsfbox{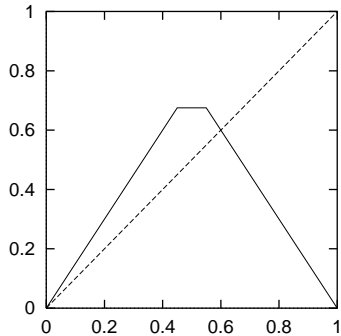}}
 \caption{Trapezoid map $T_{(a,b)}(x)$}
 \label{fig:1}
\end{figure}

The symmetric trapezoid map $x_{n+1} = T_{(a,b)}(x_n)$ 
defined in $[0, 1]$ is given by
\begin{equation}
T_{(a,b)}(x) =
\left\{
\begin{array}{ll}
 ax  & \mbox{for  $  0 \le x \le (1-b)/2 $}, \\
 a (1-b)/2 & \mbox{ for $(1-b)/2 \le x \le (1+b)/2 $ }, \\
 a(1-x)  & \mbox{ for $(1+b)/2 \le x \le 1 $ },
\end{array}
\right. 
\end{equation}
where $0<b<1$ and $a>0$. See Fig.1.\\
As $a$ is increased from 1 with fixed $b$, 
the successive period doubling bifurcations take place. 
We assign the symbol $L, C$ and $R$ to an orbit $x_i$ 
according to the following rule;
$ L\;\; {\rm for }\; x_i \in [0,(1-b)/2] \equiv I_L,
\;\; C\;\; {\rm for }\; x_i \in ((1-b)/2, (1+b)/2) 
\equiv I_C,
\;\; R\;\; {\rm for }\; x_i \in [(1+b)/2, 1] \equiv I_R$.
We denote this correspondence by $H(x_i)$.  Further,
we define the conjugate of $R$ or $L$ 
as $\bar{R}=L$ or $\bar{L}=R$, respectively.
Then, in the previous paper\cite{Ue 98} we
proved that as a slope $a$ of the trapezoid map
is increased, the period doubling bifurcation 
cascades and the symbolic sequence is 
the Metropolis-Stein-Stein sequence $R^{*m}$ 
as usual.  And,
defining $a_m$ as the onset of a $2^m$-cycle, 
we calculated $ \delta_m \equiv 
\frac{a_{m+1} -a_{m}}{a_{m+2} -a_{m+1}}$
and proved that $ \delta_m$ scales as 
\[  \delta_m \simeq a_c^{2^m}\]
where $a_c$ is the accumulation 
point of the period doubling cascade.
We called this phenomena the super-convergent period
doubling cascade.\\
Further,  we investigated the asymmetric trapezoid map
$x_{n+1} = A_{(a,b,\gamma)}(x_n)$ defined in [0,1],
\begin{equation}
A_{(a,b,\gamma)}(x) =
\left\{
\begin{array}{ll}
 ax  & \mbox{for  $  0 \le x \le \alpha $}, \\
 \alpha a & \mbox{ for $\alpha \le x \le \beta $ }, \\
 \gamma a(1-x)  & \mbox{ for $\beta \le x \le 1 $ },
\end{array}
\right. 
\end{equation}
where $\gamma$ is the ratio of two slopes of the trapezoid,
 $\alpha = \frac{\gamma}{1+\gamma} (1-b)$ and
$ \beta = \alpha +b= \frac{b+\gamma}{1+\gamma}$.
When $\gamma = 1$, $ A_{(a,b,\gamma)}
(x)$ is reduced to  $T_{(a,b)}(x)$.
In the asymmetric map, when $\gamma$ 
and $b$ are fixed
and $a$ is increased, we proved similar results
as in the symmetric map, in particular, as
the scaling of $ \delta_m$ we obtained
\[
\delta_m \simeq  \gamma ^{(-1)^m/3 } 
(a_c \gamma ^{2/3}) ^{2^m}.
\]
Finally, we gave
approximate expressions for the accumulation point
as functions of $b$ which is the length 
of the smaller side of the trapezoid
in both symmetric and asymmetric cases.  
\par

In this paper, 
we give the detailed and complete description of 
the proofs given in the previous paper.
New results in this paper are the
scaling relations for the period
doubling bifurcation starting with 
a period $p( \ge 3)$ solution which appears by
a tangent bifurcation.\par
In the following section, we treat the symmetric 
 trapezoid map, and then in the section 3, 
we treat the asymmetric map.
In sections 4 and 5,
we study the period doubling bifurcation
for period $p(\ge 3)$ solution in
symmetric and asymmetric cases, respectively.  
We give summary and discussion in the last section.\\

\section{The symmetric case}

We consider $T_{\veca }(x)
\equiv T_{(a,b)}(x)$.
For brevity we define $\veca =(a, b)$.
Let $x_M $ be the maximum value of $T_{\veca}(x)$,
 i.e., $x_M \equiv a \frac{1-b}{2}$.
Let $x_{0,1}$ be the non-zero fixed point
of $T_{\veca} (x)$.
Then the following lemma is easily proved.\\
\par
{\bf lemma 1}

For $1<a<a_1$, $T_{\veca}(x)$ has the stable fixed
point $x_{0,1} =x_M$,
which satisfies $\frac{1-b}{2}<x_M<\frac{1+b}{2}$.
$a_1$ is defined by the equation $x_M = \frac{1+b}{2}$,
that is $a_1 = \frac{1+b}{1-b}$.\par
For $a>a_1$, 
$T_{\veca} (x)$ has the unstable
fixed point $x_{0,1} =x^* \equiv \frac{a}{a+1}$
and $x^*> \frac{1+b}{2}$.  At $a =a_1, x_M =x^*.$ \\

See Fig.2.\\

\begin{figure}
     \centerline{\epsfbox{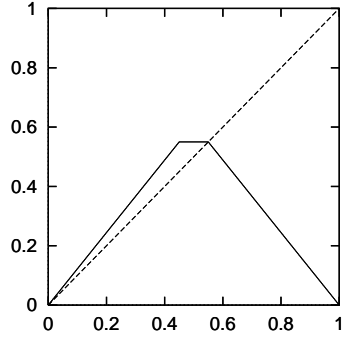}}
 \caption{Trapezoid map $T_{(a,b)}(x)$ for $a=a_1$ 
with $b=0.1$}
 \label{fig:2}
\end{figure}

In the region  $a>a_1$, by iterating $T_{\veca}(x)$ twice
and rescaling $x$ such that the domain becomes $[0, 1]$,
 we obtain a trapezoid map with
different parameter $\veca^{(1)}, 
\veca^{(1)}=(a^{(1)}, b^{(1)})$. \\

\begin{figure}[htb]
     \parbox{\halftext}{
     \epsfbox{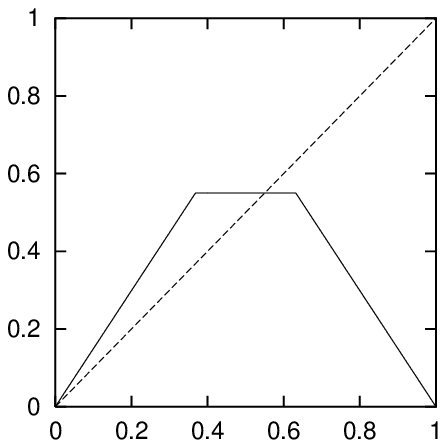}
 \caption{Trapezoid map $T^2_{(a,b)}(x)$ for $a=a_1$
with $b=0.1$.}}
 \label{fig:3}
\hspace{8mm}
     \parbox{\halftext}{
     \epsfbox{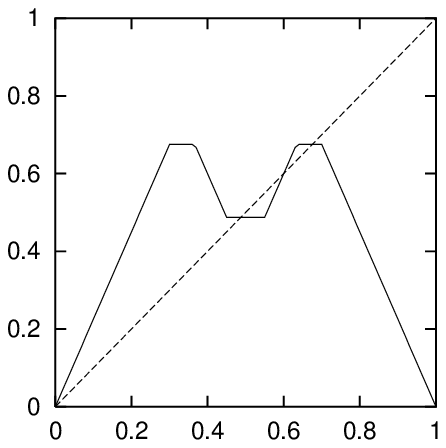}
 \caption{ Trapezoid map $T^2_{(a,b)}(x)$ for
$a_1<a<a_2$ with $b=0.1$.}}
 \label{fig:4}
\end{figure}

See Fig.3 and 4.  
In fact, we obtain the following lemma.\\
\par
{\bf lemma 2}

When  $T_{\veca} (x)$ has the non-zero unstable
fixed point $x^* = \frac{a}{a+1}(>\frac{1}{2})$,
 that is, for $a > a_1$,
by rescaling the coordinate $x$ as
$x^{(1)}=\frac{x^* -x}{2 x^* -1}$
and restricting $x$ to the interval 
$[1-x^*, x^*]$,
$T_{\veca}^2 (x)$ is transformed to 
$T_{\veca ^{(1)}} (x^{(1)})$
which is defined for $x^{(1)}$ in $[0, 1]$.
Here $\veca ^{(1)}\equiv \vecphi (\veca)
\equiv (a^2, u(a)b)$ and $u(a)=\frac{a+1}{a-1}$.\\
\par

The proof is straightforward.
$\vecphi (\veca)$ is defined as long as $a \ne 1$.
In this paper, we restrict ourselves to
the  case of  $a>1$.

Now, we define the $m$-th iteration of $\vecphi$.
That is,
\begin{eqnarray}
\veca ^{(m)} & \equiv & (a^{(m)}, b^{(m)})
 \equiv  \vecphi ^m (\veca )
=(a^{2^m}, u_m (a) b),\\
 u_m(a) & \equiv & \prod _{l=0}^{m-1} u(a^{2^l}) =
\prod _{l=0}^{m-1} \frac{a^{2^l} + 1}{a^{2^l} -  1}.
\nonumber
\end{eqnarray}
These are defined as long as $a \ne 1$.
For $a>1$, as functions of $a$,
$a ^{(m)}$ is a continuous strictly 
increasing function and 
$b ^{(m)}$ is a continuous strictly decreasing function.
$\lim_{a  \rightarrow \infty} b^{(m)}= b<1$
for any $m>0$.
Let us prove the following lemma.\\
\par
{\bf lemma 3}

For any positive integer $m$, there exists
 the unique value of $a=a_m$ such that
\begin{equation}
b ^{(m)}(a_m)=1. 
\end{equation}
   $\{a_m \}_{m=1} ^{\infty}$
is an increasing sequence, i.e.,
$1<a_1<a_2<\cdots$.  
Further, the relation $b<b ^{(m)}<1$ 
holds for  $a_m <a$ for $m \ge 1$.\\
{\bf Proof}

Let us consider the case of $m=1$.\\
$b^{(1)}(a)=1$ has the unique solution
$a=\frac{1+b}{1-b}>1$, and this is
$a_1$ defined in lemma 1.

Next, let us assume that $b ^{(m)}(a_m)=1$ and
$a_m>1$ for $m \ge 1$.
Since $b^{(m+1)}(a)=
\frac{a^{(m)}+1}{a^{(m)}-1}b^{(m)}(a)$,
$b^{(m+1)}(a_{m})>1$ follows.
Thus, there exists the unique value of $a=a_{m+1}(>
a_{m})$ such that
$b^{(m+1)}(a_{m+1})=1$.

Therefore, from the mathematical induction,
$b ^{(m)}(a)=1$ has a unique solution $a_m$
for any positive integer $m$ and
   $\{a_m \}_{m=1} ^{\infty}$
is an increasing sequence.
The inequality $b<b ^{(m)}<1$ for $a_m <a$
is immediately follows from the fact that
the function $b ^{(m)}(a)$ is strictly decreasing. 
 Q.E.D.\\

For positive integer $m$, we define   
$T_{\vecphi ^m (\veca)}(x ^{(m)})$
for $a>a_m$ from $T_{\vecphi ^{m-1} (\veca)}(x ^{(m-1)})$
successively by the same procedure as in the lemma2.
\\
We define $x ^{(m)}_M \equiv
a^{(m)} \frac{1 - b^{(m)}}{2}$, which is
the maximum value of
$T_{\vecphi ^m (\veca)}(x ^{(m)})$.
Further, we define 
$x^{(m) *}\equiv \frac{a^{(m)}}{a^{(m)}+1}$.
$x^{(0)}_M = x_M$ and $x^{(0) *} = x^*$.\\
\par
{\bf lemma 4}

For any non-negative integer $m$ and for $a_m<a$
there exists a unique non-zero fixed point for 
$T_{\vecphi ^m (\veca)}(x ^{(m)})$
which is defined in $[0,1]$.
For $a_m<a<a_{m+1}$, 
$ \frac{1 - b^{(m)}}{2}<
x^{(m)}_M < \frac{1 + b^{(m)}}{2}$ and
$x ^{(m)}_M $ is the stable 
fixed point of $T_{\vecphi ^m (\veca)}(x ^{(m)})$.
At $a=a_{m+1}$, $x ^{(m)}_M = x^{(m) *}$.
For $a_{m+1}<a$, the non-zero fixed point 
for $T_{\vecphi ^m (\veca)}(x ^{(m)})$ is 
$x^{(m) *} $
and unstable, and $x^{(m) *}> \frac{1 + b^{(m)}}{2}$.
Here, we define $x^{(0)} \equiv x,
a_0 \equiv 1, a^{(0)}\equiv a$ and
$b^{(0)} \equiv b$.\par

{\bf Proof}\\
First of all, we notice that for $m \ge 1$
$T_{\vecphi ^m (\veca)}(x ^{(m)})$ is really
a trapezoid map
because from lemma 3  $b<b^{(m)}<1$ for $a_m<a$
and for $m\ge 1$. 

For the case $m=0$, the statement follows from 
lemma 1.\par

Let us assume
that the statement holds for $m\ge 0$.
Since $x^{(m) *}=\frac{a^{(m)}}{a^{(m)}+1}$
is the unstable fixed point of
 $T_{\veca ^{(m)}} (x^{(m)})$ for $a_{m+1}<a$,
from lemma 2, taking $T_{\veca ^{(m)}}^2 (x^{(m)})$
and rescaling $x^{(m) }$ as 
$x^{(m+1)}=\frac{x^{(m) *} - x^{(m)}}{2x^{(m) *} -1}$,
we obtain $T_{\veca ^{(m+1)}} (x^{(m+1)})$
which is defined in $[0, 1]$.
The following equivalence relations are easily
proved for $m\ge0$,
\begin{equation}
b^{(m+1)}(a)=1 \Longleftrightarrow 
x ^{(m)}_M = \frac{1 + b^{(m)}}{2}
\Longleftrightarrow 
a^{(m)}= \frac{1 + b^{(m)}}{1 - b^{(m)}}.
\end{equation}
Let us consider $x ^{(m+1)}_M =
a^{(m+1)} \frac{1 - b^{(m+1)}}{2}$.
From the  relation (5), since
$b^{(m+2)}(a_{m+2})=1$ it follows that
$x^{(m+1)}_M = \frac{1 + b^{(m+1)}}{2}$ at
$a=a_{m+2}$.
Since $x^{(m+1)}_M$
is the strictly increasing function w.r.t. $a$,
we obtain $0< \frac{1 - b^{(m+1)}}{2}<
x^{(m+1)}_M < \frac{1 + b^{(m+1)}}{2}$
for $a_{m+1}<a<a_{m+2}$.  
This implies $x^{(m+1)}_M$ is the unique
non-zero fixed point
for $T_{\vecphi ^{m+1} (\veca)}(x ^{(m+1)})$
and stable.
It is easily shown that 
for $a_{m+2}<a$,  $x^{(m+1)}_M$ is no more fixed point
but $x^{(m+1) *}= \frac{a^{(m+1)}}{a^{(m+1)}+1}$
becomes unstable fixed point
and $x^{(m+1) *}> \frac{1 + b^{(m+1)}}{2}$.
At $a=a_{m+2}, a^{(m+1)}_M= x^{(m+1) *}$ holds.
Thus, by the mathematical induction, the proof completes.\\
\par
{\bf lemma 5}

The unique non-zero fixed point of 
$T_{\vecphi ^m (\veca)}(x ^{(m)})$ in lemma 4
is a member of a periodic cycle with prime period $2^m$
of $T_{\veca}(x)$.

{\bf Proof}

The case of $m=0$, it is trivial.
Now, let us fix $m(\ge 0)$ and assume that
for $0\le k \le m$ the unique non-zero fixed point of 
$T_{\vecphi ^k (\veca)}(x ^{(k)})$ 
which exists for $a_k<a$ is the member of
the cycle with prime period $2^k$.
Thus, from lemma 4 for $a_{m+1}<a$
the unique non-zero fixed point of 
$T_{\vecphi ^{m+1} (\veca)}(x ^{(m+1)})$
does not correspond to any periodic points
with period less than $2^{m+1}$ because it
is stable for $a_{m+1}<a<a_{m+2}$.
Thus, the non-zero fixed point of 
$T_{\vecphi ^{m+1} (\veca)}(x ^{(m+1)})$
is the member of the cycle
with prime period  $2^{m+1}$.
This completes the proof.\\
\par
Thus, we obtain the following theorem.\\
\par
{\bf Theorem 1}

$x=0$ is period 1 solution of 
$T_{\veca}(x)$ for any $a>0$.  Except for this,
for any non-negative integer $m$, 
for $a_m<a<a_{m+1}$ there is 
the unique cycle with prime period  $2^k$ for $k=0,1,
\cdots,m$ and no other periodic 
points exist.  The periodic solution with
prime period $2^m$ is stable for
$a_m <a <a_{m+1}$ and unstable for
$a_{m+1}<a$.
At $a=a_{m+1}$, the period $2^m$ cycle and
the period $2^{m+1}$ cycle coincide.\\
{\bf Proof}

For the case of $m=0$, the statement is trivial.
Let us assume that the statement holds for
 $0,1,\cdots, m$.  Then
we only have to prove that for $a_{m+1}<a<a_{m+2}$,
except for the period $2^k$ cycles for $k = 1 \sim m+1$,
there is no other cycle.
Let us consider any initial point $x$ in (0, 1) 
which does not corresponds to the 
unstable cycles of period $2^k, (k=0 \sim m).$ 
Then, the orbit $T^{i}_{\veca}(x)$
 finally enters the domain of the map 
$T_{\vecphi ^m (\veca)}(x ^{(m)})$
if $i$ is appropriately chosen.
Thus, since any point except for $x ^{(m)}=0$ and 1
converges to the fixed point of 
$T_{\vecphi ^m (\veca)}(x ^{(m)})$ for  
$a_{m+1}<a<a_{m+2}$,
the nonexistence of other periodic points
for $a_{m+1}<a<a_{m+2}$ follows.
The latter part of the theorem follows from lemma 4
immediately.\\

From the above arguments we
conclude that as $a$ is increased from 1, the 
period doubling bifurcation cascades,
and at $a=a_m$ the periodic solution with 
prime period $2^m$ appears for $m\ge 0$.\par

Now, let us investigate the upper bound of the 
sequence $\{ a_m \}$.
$x^{(m)}_M(a)= a^{(m)}\frac{1-b^{(m)}}{2}$ 
is defined for $ a>1$ and the 
strictly increasing
function w.r.t $a$.  
For any $m >0, x^{(m)}_M(a_m)=0$ and
$\lim_{a \rightarrow \infty} x^{(m)}_M(a)=\infty$
follows.
Thus, for $m>0$ we define $a_M(m)$ as the
unique solution of the equation 
\begin{equation}
 x^{(m)}_M(a)=1.
\end{equation}
For $m=0$, we define 
$a_M(0)=\frac{2}{1-b}$.  We abbreviate this as $a_M$.
Then, we obtain the following lemma.\\
\par
{\bf lemma 6}

\begin{eqnarray}
&& 1<a_1<a_2<\cdots<a_m<a_{m+1}<\cdots \\
&& <a_M(m+1)<a_M(m)\cdots<a_M (2)<a_M (1)<a_M. \nonumber
\end{eqnarray}

{\bf Proof}\\
For $m>0$, since $x_M ^{(m)}(a_m)=0,  a_m<a_M(m)$ follows.
For $m \ge 0$, $x_M ^{(m+1)}(a)$ is rewritten as
\begin{equation}
 x_M ^{(m+1)}(a)=\{a^{(m)} \}^2 
\frac{1}{ a^{(m)}-1} ( x_M ^{(m)}(a)-
\frac{1+b^{(m)} }{2}).
\end{equation}
Then, $x_M ^{(m+1)}(a_M(m))=\frac{ a^{(m)}}
{ a^{(m)}-1} >1.$
Therefore, $a_M(m+1)< a_M(m)$ for $m \ge 0$. 
Thus, we obtain 
$a_m<a_{m+1}<a_M(m+1)<a_M(m)$  for $m\ge 0$.
Q.E.D. \\

From this, 
\begin{equation}
 a_c \equiv \lim_{m \rightarrow 
\infty}a_m \le \lim_{m \rightarrow\infty}a_M(m)<a_M 
\end{equation}
follows.  Therefore $a_c$ is finite.

Now, let us investigate the symbolic sequence.
The sequence of $R$ and $L$
for the period $2^m$ solution becomes
Metropolis-Stein-Stein(MSS) sequence.  
Let us prove this.

For $m \ge 0$, we define the onset of the $2^n$-cycle 
for $T_{\vecphi ^m (\veca)}(x^{(m)})$
as $a^{(m)}_n$ and
the $i$-th orbit of the period $2^n$ cycle
 as $x^{(m)} _{n,i}$.
$a^{(0)} _n = a_n$.  
For $m=0$, we often omit the superscript $(0)$.
As $x_{m,1}$ we set the largest $x$ value
among $2^m$ members of the periodic cycle.
That is, $x_{m,1}=x_M=a(1-b)/2$ when the cycle is stable.
Further, for $m\ge 0$ 
we divide the coordinate space $[0,1]$ of  $x^{(m)}$ 
into the three intervals as
$I^{(m)}_L \equiv [0, \frac{1-b^{(m)}}{2}],
I^{(m)}_C \equiv (\frac{1-b^{(m)}}{2},
\frac{1+b^{(m)}}{2})$ and
$I^{(m)}_R \equiv [\frac{1+b^{(m)}}{2},1].$\\
\par
{\bf lemma 7}

For any positive integer $m$, when $a_m < a$,
\begin{enumerate}
\item
 $I^{(m)}_C \ni x^{(m)}$ is equivalent to 
 $I^{(m-1)}_C \ni x^{(m-1)}$,
\item
if   $I^{(m)}_L \ni x^{(m)}$ then
 $I^{(m-1)}_R \ni x^{(m-1)}$,
\item
if   $I^{(m)}_R \ni x^{(m)}$ then
 $I^{(m-1)}_L \ni x^{(m-1)}$,
\end{enumerate}
where $x^{(m)}$ and $x^{(m-1)}$ are related by the
coordination transformation used to define 
$T_{\vecphi ^m (\veca)}(x^{(m)})$ from
$T_{\vecphi ^{m-1} (\veca)}(x^{(m-1)})$.

{\bf proof}

For $a_m < a$, $x^{(m)}=0, \frac{1-b^{(m)}}{2},
\frac{1+b^{(m)}}{2}$ and 1 correspond to 
$x^{(m-1)}= x^{(m-1) *}, \frac{1+b^{(m-1)}}{2},
\frac{1-b^{(m-1)}}{2}$ and $1-x^{(m-1) *}$,
respectively.  Since this correspondence is linear, the
statements hold.\\
\par
{\bf lemma 8}

The fixed point $x^{(m)} _{0,1}$ of 
$T_{\vecphi ^m (\veca)}(x^{(m)})$
corresponds to $x_{m,2^m}$. 

{\bf proof}

Let $x_{m,i}$ be the orbit
of $2^m$ solution corresponding to
$x^{(m)} _{0,1}$.
From lemma 4, for $a_m<a<a_{m+1}, 
x^{(m)} _{0,1}=x^{(m)} _M$ and
  $I^{(m)}_C \ni x^{(m)} _{0,1}$.
Then, from lemma 7,   $I_C \ni x_{m,i}$ follows.
Therefore, $x_{m,i}$ is mapped to $x_M$ by
$T_{\veca}$, which is $x_{m,1}$.
Thus, $i=2^m$.\\
\par
{\bf lemma 9}

For any positive integer $m$, symbols
$H(x_{m,i})$ for the number of $2^m$ cycle
satisfy the followings.

\begin{enumerate}
\item
$H(x_{m,i})$ does not change for 
$a_m \le a$ and for $i=1,\cdots,2^m -1$.

\item
For even $m$,
\[ H(x_{m,2^m}) =
\left\{
\begin{array}{ll}
 L  & \mbox{for  $  a=a_m $}, \\
 C  & \mbox{ for $  a_m < a < a_{m+1} $ }, \\
 R  & \mbox{ for $  a_{m+1} \le a $ }.
\end{array}
\right. \]
and for odd $m$,
\[ H(x_{m,2^m}) =
\left\{
\begin{array}{ll}
 R  & \mbox{for  $  a=a_m $}, \\
 C  & \mbox{ for $  a_m < a < a_{m+1} $ }, \\
 L  & \mbox{ for $  a_{m+1} \le a $ }.
\end{array}
\right. \]
\item
$\{ H(x_{m,i}) \}$ is the MSS sequence 
$R^{*m}$
\footnote{ Strictly speaking, 
the last symbol of the sequence
in our definition is $R$ or $L$ and is different from that
in the  MSS sequence, $C$} for $a_{m+1} \le a $.
\end{enumerate}

{\bf proof}

Let us consider the case of $m=1$.
At $  a=a_1 $, 
$x _{0,1} = \frac{1+b}{2} \in I_R$.
Then, at $  a=a_1, H(x_{0,1})=
H(x_{1,1})=H(x_{1,2})=R $.
For $  a_1 < a < a_2 $, 
$x^{(1)} _{0,1} =x^{(1)} _M= \frac{a^{(1)}(1-b^{(1)})}{2}$
and for $  a_2 \le  a , x^{(1)} _{0,1} =x^{(1) *}$.
From lemma 8, $x^{(1)} _{0,1}$ corresponds to
$x_{1,2}$.  Thus, from lemma 7,
$ H(x_{1,2}) $ is $C$ for $  a_1 < a < a_2 $, 
and is $L$ for  $  a_2 \le  a $.
On the other hand, for $a_1 \le a$,
since $ x_{1,1}\ge x_{0,1}= \frac{a}{a+1}$ 
then $ H(x_{1,1})=R $.
Therefore, $H(x_{1,1})H(x_{1,2})=RL$ for
$a_2 \le a$.  Therefore, for $m=1$, the
lemma holds.

Next, we assume that the lemma holds for the case
of $m(\ge1)$.  At $  a=a_{m+1}$ , 
$x_{m+1,i}$ and $x_{m+1,i+2^m}$ emerge from
$x_{m,i} (i=1 \sim 2^m)$.
Then, at $  a=a_{m+1}$,
$H(x_{m,i})=H(x_{m+1,i})=H(x_{m+1,i+2^m})$ 
for $i=1,\cdots, 2^m$.  These are $L$ or $R$.
Let us assume that at some value of $a(>a_{m+1})$,
$H(x_{m+1,i})$ becomes $C$.
Then, $x_{m+1,i}$ is mapped to $x_M$
 by $T_{\veca}$.  Thus, in this case, $a<a_{m+2}$
and $x_{m+1,i+1}=x_M=x_{m+1,1}$ follow.
  So, $i$ should be $2^{m+1}$.
Thus, $x_{m+1,i}(i=1 \sim 2^{m+1}-1)$
does not change its symbol for $a_{m+1}\le a$.
From lemma 8 $x_{m+1,2^{m+1}}$ corresponds
to $x^{(m+1)}_{0,1} = x^{(m+1)}_M$ for
$  a_{m+1}\le a \le a_{m+2}$.
Thus, $H(x_{m+1,2^{m+1}})=C$ for
$  a_{m+1}< a < a_{m+2}$.
For $  a_{m+2}\le a$, 
$x^{(m+1)}_{0,1} = x^{(m+1) *} \in I^{(m+1)}_R$.
Thus, from lemma 7, for $  a_{m+2}\le a$, 
\[ H(x_{m+1,2^{m+1}})=
\left\{
\begin{array}{ll}
 L  & \mbox{ for  odd $  m+1 $}, \\
 R  & \mbox{ for  even $ m+1 $}.
\end{array}
\right. \]
Therefore, the statements 1 and 2 hold for $m+1$.\\
At $a=a_{m+1}, H(x_{m+1,2^{m+1}})=H(x_{m,2^m})$.
Then, from the assumption,
\[ H(x_{m+1,2^{m+1}})=
\left\{
\begin{array}{ll}
 R  & \mbox{ for  odd $  m+1 $}, \\
 L  & \mbox{ for  even $ m+1 $}.
\end{array}
\right. \]

Let $H(x_{m,1})H(x_{m,2})\cdots H(x_{m,2^m})$
be the MSS sequence for $a \ge a_{m+1}$.
From the above argument, 
for $a \ge a_{m+2}$, the sequence for
$2^{m+1}$ cycle is
\begin{eqnarray*}
&& H(x_{m,1})H(x_{m,2})\cdots H(x_{m,2^m-1})H(x_{m,2^m})\\
&& \times H(x_{m,1})H(x_{m,2})
\cdots H(x_{m,2^m-1}) \overline{H(x_{m,2^m})}.
\end{eqnarray*}
This is the MSS sequence.
Thus, the statement 3 is proved.
This completes the proof.\\

Now, we derive the equations for which
$a_m$ and $a_c$ should satisfy, respectively,
and obtain the asymptotic expression for $\delta_m$.
First, we assign 0 or 1 to any orbit $x_i$ 
with the symbol $L$ or $R$, respectively.
We denote this correspondence as $s_i=s(x_i)$.
Further, we define the function 
$f_s (x)$ for $x \in I_L \cup I_R$ as
\begin{equation} 
f_s (x) = sa +(1-2s)ax,  
\end{equation}
where $s=s(x)$.
Starting from the maximum value of 
$T_{a,b}(x)$, $x_1=x_M= a(1-b)/2$,
if $ x_1,  x_2,\cdots ,x_{n-1} \in  I_L \cup I_R$,
$x_n$ is expressed as
\begin{eqnarray}
x_n &=& f_{s_{n-1}}(x_{n-1})= 
f_{s_{n-1}}\circ f_{s_{n-2}}\circ
\cdots \circ f_{s_1}(x_1)= \sum_{l=1}^n \xi _l a^l, \\
\xi_l &=& s_{n-l} \prod_{j=n-l+1}^{n-1}(1-2s_j)
 \;\; {\rm for }
\;\; 2 \le l \le n-1, \nonumber \\
\xi_1 &=& s_{n-1},\;\;
\xi_n = \frac{1-b}{2} \prod_{j=1}^{n-1}(1-2s_j). \nonumber
\end{eqnarray}
For $m \ge 1$ let us consider the period $2^m$ 
solution $\{x_{m,i}\}$
for $a_m \le a \le a_{m+1}$.
\begin{eqnarray}
x_{m,2^m} &=& f_{s_{2^m -1}} (x_{m, 2^m -1})
 \equiv \sum_{l=0} ^{2^m -1} c_l ^{(m)} 
a ^{2^m -l}\equiv F_m (a),\\
c_l ^{(m)} &=& s_{l} \prod_{j=l+1}^{2^m -1}(1-2s_j)
=\xi_{2^m-l},\;\; {\rm for }
\;\; 1 \le l \le 2^m -2, \nonumber \\
c_0 ^{(m)} &=& \frac{1-b}{2} 
\prod_{j=1}^{2^m -1}(1-2s_j)
=s_0\prod_{j=1}^{2^m -1}(1-2s_j),\nonumber \\
c_{2^m -1} ^{(m)} &=& s_{2^m-1},\nonumber
\end{eqnarray}
where $s_0 \equiv \frac{1-b}{2}$.
Note that $ F_m (a)$ is 
determined by the sequence $( s_1, \cdots, s_{2^m-1})$
and is independent of $s_{2^m}$.
From lemma 9,  the symbols $H(x_{m,i})(i=1 \sim
2^m-1)$ do not change for $a_m \le a$.  Further, 
for any positive
integer $j$ these symbols are equal to the first
$2^m-1$ symbols of the period $2^{m+j}$ solution
 when $a_{m+j} \le a$.
Therefore, $ F_m (a)$ expresses 
$x_{2^{m+j},2^m}$ in the region $ [a_{m+j},a_{m+j+1}]$.
From the statement 2 in
lemma 9, the following relations follow
\begin{eqnarray}
 x_{m,2^m}(a_m)& = & \frac{ 1+(-1)^{m-1} b}{2},\\
 x_{m,2^m}(a_{m+1})& = & \frac{ 1+(-1)^m b}{2}.
\end{eqnarray}
  Thus, we obtain the following conditions
for $a_m$,
\begin{equation}
 F_m(a_m)  =  \frac{ 1+(-1)^{m-1} b}{2},
\end{equation}
or
\begin{equation}
 F_{m-1}(a_m)  =  \frac{ 1+(-1)^{m-1} b}{2}.
\end{equation}
Since the symbolic sequence is the Metropolis-Stein-Stein
sequence $R^{*m}$, 
\begin{eqnarray}
\Pi_{j=1}^{2^m-1}(1-2s_j)&=&(-1)^m,\\
s_{2^m}&=&\frac{1+(-1)^m}{2}
\end{eqnarray}
 follow.  From these relations, for $l \ge 0$ we
obtain
\[r_l \equiv 
\{ \prod_{j=1}^{2^m -1}(1-2s_j) \}^{-1}c_l ^{(m)}
=(-1)^m c_l ^{(m)}=s_l  \prod_{j=1}^{l}(1-2s_j).
\]
That is, $r_l$ is $m$-independent as
long as it is defined.
Thus, we get
\begin{eqnarray}
r_l&=&s_l \prod_{j=1}^{l}(1-2s_j)
 \mbox{ for any } l (\ge 0),\\
r_0 & =& (1-b)/2,\;\; r_{2^m}= -( 1+(-1)^m)/2
 \mbox{ for any } m (\ge 0),
\end{eqnarray}
and for $l>0$ the successive values 
$(r_l, r_{l+1})$ take  the following
six sets of values, $(0,\pm 1),(\pm 1,0),(1,-1),(-1,1)$.
\footnote{In paper \cite{Ue 98}, two cases
$(0, -1)$ and $(-1, 0)$ are missing.}
Let us define
\begin{equation}
G_m (a) \equiv (-1)^m a^{-2^m} F_m(a)=
 \sum_{l=0} ^{2^m -1} r_l  a ^{-l}.
\end{equation}
\begin{equation}
G_{\infty}(z) \equiv \lim_{m \rightarrow \infty} G_m (z).
\end{equation}
Then, $G_{\infty}(z)$ is the analytic function for $|z|>1$.
The equation (2$\cdot$13) becomes
\begin{equation}
 G_m(a_m)=a_m ^{-2^m} \frac{ (-1)^m -b}{2},
\end{equation}
and the accumulation point $a_c$ satisfies the equation,
\begin{equation}
G_{\infty} (a_c)=0.
\end{equation}
Let us estimate $a_m$ for large $m$.  
Putting $a_m = a_c - \epsilon _m$
and using the mean value theorem,
the right hand side of eq.(2$\cdot$21) is rewritten as
\[
a_m ^{-2^m} \frac{ (-1)^m -b}{2}
= ( a_c ^{-2^m}+2^m \hat{a}_m ^{-2^m-1} \epsilon _m )
  \frac{ (-1)^m -b}{2},\]
where $a_m < \hat{a}_m <a_c$.
On the other hand, the left hand side of eq.(2$\cdot$21) is
expressed as
\[G_m(a_m)=G_m(a_c)-G_m'(\bar{a}_m)\epsilon _m,\]
where $a_m < \bar{a}_m <a_c$.
Thus, we obtain from eq.(2$\cdot$22)
\begin{equation}
  \epsilon _m = \{ G_m (a_c)+
\frac{b-(-1)^m}{2}a_c ^{-2^m} \}
(1+ \bar{h}_m)/G'_{\infty} (a_c)
\end{equation}
where $\bar{h}_m=G'_{\infty} (a_c)/(G_m'(\bar{a}_m)+
2^m (\hat{a}_m)^{-2^m -1} \frac{(-1)^m -b}{2} )-1$
 and $\lim_{m \rightarrow \infty} \bar{h}_m=0$.
Here, we assume $G'_{\infty} (a_c)\ne 0$,
which is proved later.
Equation ($2\cdot22$) is rewritten as
\begin{equation}
G_{\infty} (a_c)=  G_m(a_c) + \sum _{l=0} ^{\infty}
r_{2^m + l} a_c ^{ - l - 2^m} = 0.
\end{equation}
Using the relation
\begin{equation}
r_{2^m +l} =-r_l\;\; \mbox{\rm  for } 1\le l \le 2^m -1, 
\end{equation}
eq.($2\cdot24$) is further changed to the following.
\begin{equation}
G_{\infty} (a_c)=  G_m(a_c)(1-  a_c ^{- 2^m}) 
+ a_c ^{- 2^m}(r_0 + r_ {2^m}
+ a_c ^{- 2^m} \sum _{l=0} ^{\infty}r_{2^{m+1} + l}
 a_c ^{ - l}) = 0.
\end{equation}
Thus, we get
\begin{equation}
 G_m(a_c) = \frac{b+ (-1) ^m}{2} a_c ^{- 2^m} + 
(a_c ^{- 2^m} )^2 q_m ,
\end{equation}
where $ q_m=  \frac{1}{1-a_c ^{- 2^m}}
( \frac{b+ (-1) ^m}{2} - 
\sum_{l=0}^{\infty}r_{2^{m+1} + l}a_c ^{ - l})$
 and $|q_m|<\frac{a_c(2a_c-1)}{(a_c-1)^2}$.
Substituting eq.(2$\cdot$27) into eq.(2$\cdot$23) we obtain,
\begin{equation}
  \epsilon _m = \frac{b a_c ^{-2^m}}{G'_{\infty} (a_c)} 
(1 + h_m),
\end{equation}
where $h_m=\bar{h}_m (1+ \frac{q_m}{b}a_c ^{- 2^m})
+\frac{q_m}{b} a_c ^{- 2^m}$ and
$\lim_{m \rightarrow \infty} h_m=0$.
Thus,
\begin{eqnarray}
 \delta_m&=&\frac{\epsilon_m-\epsilon_{m+1}}
{\epsilon_{m+1}-\epsilon_{m+2}}
= a_c ^{2^m}(1+l_m),\\
&& l_m=[ h_m -h_{m+1}-a_c ^{- 2^m}
\{1+h_{m+1}- a_c ^{- 2^m}(1+h_{m+2})\}] \nonumber\\
&& /[ 1+h_{m+1}- a_c ^{- 2^{m+1}}(1+h_{m+2})],\nonumber
\end{eqnarray}
and $\lim_{m \rightarrow \infty} l_m=0$.

Next, we give the alternative relation
for the onset point $a_m$.  As is
shown in lemma 3, at $a=a_m$ 
we have the following equation (2$\cdot$2) for $m >0$,
\[ b^{(m)}(a_m)=u_m(a_m)b=1.\]
Defining $v(a)\equiv \frac{a-1}{a+1}=
1/u(a)$ and $v_m(a) \equiv
1/u_m(a)
=\prod_{l=0}^{m-1}v(a ^ {2^l})$,
we obtain for $m>0$
\begin{equation}
 v_m(a_m)=b.
\end{equation}
For $m \ge  1$, $v_m(a)$ is rewritten as
\begin{eqnarray}
 v_m(a)& =& \frac{1-a^{-1}}{1+a^{-2^{m-1}}}
\prod_{l=0} ^{m-2} (1-a^{-2^l}),\\
&& \prod_{l=0} ^{-1} (1-a^{-2^l}) \equiv 1,
\;\; v_1(a)=v(a).\nonumber
\end{eqnarray}
As is easily shown, for $m \ge 2 $, $G_{m-1}(a)$ is
expressed by $ v_m(a)$ as follows,
\begin{equation}
G_{m-1}(a) = \frac{1}{2} 
\{ (1+a^{-2^{m-1}}) v_m(a)-b-(-1)^m a^{-2^{m-1}} \}.
\end{equation}
See Appendix A.
Then the equation obtained from the equation ($2\cdot14$)
\begin{equation}
G_{m-1}(a_m) = (-1)^{m-1} a_m ^{-2^{m-1}}
\frac{1 +(-1)^{m-1}b}{2} 
\end{equation}
is equivalent to the eq.(2$\cdot$30). 
From the relation (2$\cdot$32),
\begin{equation}
G_{\infty}(a) = \frac{1}{2} 
( v_{\infty}(a) -b)
\end{equation}
follows for $a>1$.  Then $G_{\infty}(a_c)=0$
 is equivalent to
$v_{\infty}(a_c)=b$ which 
follows from eq.(2$\cdot$30) immediately.  
Since 
\begin{equation}
 v'_{\infty}(a) = 2 v_{\infty}(a) 
\sum _{l=0} ^{\infty}
 \frac{2^l a^{2^l -1}}{ a^{2^{l+1}}-1},
\end{equation}
we obtain
\begin{equation} 
v_{\infty}'(a)>0, \;\;
G_{\infty}'(a)>0\;\; \mbox{\rm  for } a>1.
\end{equation}
Putting $a=a_c$ in the eq.(2$\cdot$35), we get
\begin{eqnarray}
 v'_{\infty}(a_c)& = & 2 b  \tau(a_c),\\
&&\tau(a_c) \equiv
\sum _{l=0} ^{\infty}
 \frac{2^l a_c ^{2^l -1}}{ a_c ^{2^{l+1}}-1}>0.
\end{eqnarray}
Thus,
\begin{equation} G'_{\infty}(a_c)=
v'_{\infty}(a_c)/2=b\tau(a_c)>0.
\end{equation}
Therefore, substituting this expression(2$\cdot$38) 
into eq.(2$\cdot$28)
we obtain 
\begin{equation}
 \epsilon_m =\frac{ a_c  ^{-2^m}}{ \tau(a_c)}
(1+h_m).
\end{equation}
\par

\section{ The asymmetric case}
\noindent

As in the symmetric case, we can discuss
the period doubling cascade 
when $a$ is increased with fixed $b$ and $\gamma$ 
in the asymmetric case. 
For brevity we define $\veca =(a, b, \gamma)$.
$\alpha$ and $\beta$ are expressed as
\[ \alpha=\frac{\gamma (1-b)}{1+\gamma},\;\;
 \beta= \alpha +b = \frac{b+\gamma}{1+\gamma}.\]
$\alpha, \beta$ and $\gamma$ are related by
$\alpha=\gamma(1-\beta)$.
As long as $0<b<1$, 
it holds that $0<\alpha<1$ and $0<\beta<1$.
 Let us define $a_L\equiv a$ and $a_R \equiv \gamma a$.
Let $x_M $ be the maximum value of $A_{\veca}(x)$,
 i.e., $x_M \equiv a \alpha$.
We define  $x^* \equiv \frac{\gamma a}{1+\gamma a}$.
And let $x_{0,1}$ be the non-zero
fixed point of $A_{\veca}(x)$.
 The following lemma is easily proved.\\
\par
{\bf lemma 1'}

\begin{enumerate}
\item The case of $\beta \ge 1/2$.
For $1<a<a_1$, $A_{\veca}(x)$ has the stable fixed
point $x_{0,1} = x_M$,
which satisfies $\alpha<x_M<\beta$.
$a_1$ is defined by the equation $x_1 =\beta$,
that is $a_1= \frac{\beta}{\alpha}$.
\item The case of $\beta<1/2$.
For  $1<a<\frac{\beta}{\alpha}$,
$A_{\veca}(x)$ has the stable
fixed point  $x_{0,1} =x_M$, which
 satisfies $\alpha<x_M<\beta$.  
For $a \ge \frac{\beta}{\alpha}$
the non-zero fixed point  
becomes $x^*$ and  continues to be stable until
$\frac{1- \beta}{\alpha}$.  Thus, in this case
we define $a_1 =\frac{1- \beta}{\alpha}
=\frac{1}{\gamma}$.
\end{enumerate}

In both cases, for $a>a_1(>1)$, 
$A_{\veca} (x)$ has the unstable
fixed point $x_{0,1}=x^*>\beta $.\par
\vspace{1cm}

In the region  $a>a_1$, by iterating $A_{\veca}(x)$ twice
and rescaling $x$, we obtain a trapezoid map with
different parameter $\veca^{(1)}, 
\veca^{(1)}=(a^{(1)}, b^{(1)}, \gamma^{(1)})$. 
We obtain the following lemma.\\
\par
{\bf lemma 2'}

When  $A_{\veca} (x)$ has the non-zero unstable
fixed point of $x^* = \frac{\gamma a}{\gamma a+1}
(>\beta)$,
by rescaling the coordinate $x$ as
$x^{(1)}=\frac{x^* -x}{\Delta x}$, 
$A_{\veca}^2 (x)$ is transformed to 
$A_{\veca ^{(1)}} (x^{(1)})$
which is defined for $x^{(1)}$ in $[0, 1]$,
where  several parameters and variables
are defined as
\begin{eqnarray}
\Delta x & = & \gamma (a-1)/(1+ \gamma a),\hspace{0.5cm}
\veca ^{(1)}= \vecphi (\veca),\nonumber \\
\vecphi & :& (a, b, \gamma ) \rightarrow 
(a^{(1)}, b^{(1)}, \gamma^{(1)} ) ,\\
&& a^{(1)} = a_R ^2 = (\gamma a)^2, \;\;  
a_R ^{(1)} = a_L a_R = \gamma a^2, \nonumber \\
 b^{(1)} & = & u(a, \gamma) b,\;\;
u(a, \gamma) \equiv \frac{\gamma a +1}{ \gamma (a-1)},\;\;
\gamma^{(1)}= 1/\gamma, \;\;
\alpha ^{(1)} = \frac{\alpha  a - \beta}{ \gamma (a-1)}.
\nonumber
\end{eqnarray}
\\
\par
The proof is straightforward.
Regardless of the value of $\beta$,
$\beta^{(1)}>1/2$ follows.
Now, we define the $m$-th iteration of $\vecphi$
for $m \ge 0$.
That is,
\begin{eqnarray}
\veca ^{(m)} &  \equiv &  \vecphi ^m (\veca )
\equiv  (a^{(m)}, b^{(m)}, \gamma ^{(m)}),\\
a^{(m)} & =& \gamma^{2(2^m - (-1)^m)/3} 
a^{2^m},\hspace{0.5cm}
a_R^{(m)}  = \gamma^{(2^{m+1} + (-1)^m)/3} a^{2^m},
\nonumber \\
b^{(m)}& =& u_m(a, \gamma) b, \;\; 
 u_m(a, \gamma) \equiv { \prod _{l=0}^{m-1} 
u(a^{(l)},  \gamma^{(l)}) }, \hspace{0.5cm}
\gamma ^{(m)} = \gamma ^{(-1)^m},\nonumber \\
\alpha^{(m)} & =& \frac{\gamma^{(m)}(1-b^{(m)})}
{1+\gamma^{(m)}},\;\; 
\beta^{(m)}  = \frac{ b^{(m)}+\gamma^{(m)}}
{1+\gamma^{(m)}},\nonumber
\end{eqnarray}
where $u_0(a, \gamma)\equiv 1, a^{(0)}\equiv a,
b^{(0)}\equiv b$ and $\gamma^{(0)}\equiv \gamma$.

Since $a^{(m)}=(\gamma a)^{2^m} 
\gamma ^{-2 [2^{m-1}+(-1)^m]/3}$
 and $\gamma a_1 \ge 1$,
it is easily shown that 
$a^{(m)}>1$ holds for $a>a_1$ and $m \ge 0$.
Thus, for  $m \ge 1$, $u_m(a, \gamma)$
and $b^{(m)}(a, \gamma)$  are defined for $a>a_1$.
In the below, we assume $a>1$.
For $m \ge 1$, 
$b ^{(m)}$ is a continuous strictly decreasing function
w.r.t. $a$ for $a>a_1$, and 
 $\lim_{a \rightarrow \infty} b^{(m)}(a, \gamma)=b$.

Let us prove the following lemma.\\
\par
{\bf lemma 3'}

For integer $m\ge 2$, there exists
the unique value of $a=a_m$ greater
than 1  such that 
\begin{equation}
a^{(m-1)}= \frac{\beta^{(m-1)}}{\alpha^{(m-1)}},
\end{equation}
which is equivalent to 
\begin{equation}
b ^{(m)}(a_m)=1. 
\end{equation}
   $\{a_m \}_{m=1} ^{\infty}$
is the increasing sequence, $1<a_1<a_2<\cdots$, and
for $a_m <a, b<b ^{(m)}<1$ for $m \ge 1$.

{\bf Proof}

The following equivalence relations
are easily proved if these quantities are defined.
\begin{equation}
b^{(m)}(a)=1 \Longleftrightarrow 
x_M^{(m-1)}=\beta ^{(m-1)}
\Longleftrightarrow 
a^{(m-1)}= \frac{\beta^{(m-1)}}{\alpha^{(m-1)}}>1.
\end{equation}

Let us consider the $m=2$ case.  For $a>1$,
 $b^{(1)}$
is defined and
 $a_2$ is the solution of 
the equation 
$a^{(1)}= \frac{\beta^{(1)}}{\alpha^{(1)}}$.
There is the unique solution greater than 1
of this equation,
 $a_2 = \frac{1}{2\alpha} 
( 1+ \sqrt{ 1- \frac{4 \alpha ^2}{ \gamma}})$.
$a_2>a_1$ is easily proved.
From the relation (3$\cdot$4),
$b^{(2)}=1$ at $a=a_2$.

\noindent
Next, let us assume that
 $b^{(m)}(a_m)=1$ for $m(\ge 2)$ and $a=a_m >1$. 
Then for $a > a_m$, $b<b^{(m)}<1$ and the function
 $ \beta^{(m)}/\alpha^{(m)}= 1 +\frac{1+\gamma^{(m)}}{\gamma^{(m)}}
\frac{b^{(m)}}{1-b^{(m)}}$ is continuous and 
decreasing w.r.t. $a$.
On the other hand, $a^{(m)}$ is the
continuous increasing function w.r.t. $a$.
Further, we obtain following limits.
\begin{eqnarray*}
\lim_{a \rightarrow a_m +0}
\frac{ \beta^{(m)}}{\alpha^{(m)}}&=&\infty,
\;\; \lim_{a \rightarrow \infty}
\frac{ \beta^{(m)}}{\alpha^{(m)}}=
1+\frac{1+\gamma^{(m)}}{\gamma^{(m)}}
\frac{b}{1-b}= \mbox{ finite},\\
\lim_{a \rightarrow a_m +0}
a^{(m)}&=& \mbox{ finite},\;\;
 \lim_{a \rightarrow \infty}
a^{(m)}=\infty.
\end{eqnarray*}
Thus, there exists the unique value of $a_{m+1}(>a_m)$
such that $a^{(m)}=\beta^{(m)}/\alpha^{(m)}>1$.
Thus,
$b^{(m+1)}(a_{m+1})=1.$
Therefore, the first half of the lemma is proved.
The second half immediately follows from
the decreasing property of $b^{(m)}$
 and the facts $b^{(1)}(a=\beta /\alpha)=1,
a_1 \ge \beta/ \alpha$ and $1<a_1<a_2$. Q.E.D.\\

For positive integer $m$, we define   
$A_{\vecphi ^m (\veca)}(x ^{(m)})$
for $a>a_m$ from $A_{\vecphi ^{m-1} (\veca)}(x ^{(m-1)})$
successively by the same procedure as in the lemma2'.
\\
We define $x ^{(m)}_M \equiv
a^{(m)} \alpha^{(m)}$, which is
the maximum value of
$A_{\vecphi ^m (\veca)}(x ^{(m)})$.
Further, we define 
$x^{(m) *}\equiv \frac{\gamma^{(m)} a^{(m)}}
{\gamma^{(m)}a^{(m)}+1}$.
$x^{(0)}_M = x_M$ and $x^{(0) *} = x^*$.\\
\par
{\bf lemma 4'}

For non-negative integer $m$ and for $a_m<a$
there exists unique non-zero fixed point for 
$A_{\vecphi ^m (\veca)}(x ^{(m)})$.
For $a_m<a<a_{m+1}$, the fixed point is stable.
For $m \ge 1$, the stable fixed point is
$x ^{(m)}_M$ and $\alpha^{(m)} < x ^{(m)}_M <
\beta^{(m)}$.\footnote{For $m=0$, see lemma 1'}
At  $a=a_{m+1}$, $x ^{(m)}_M= x^{(m) *}=\beta ^{(m)}$.
For $a_{m+1}<a$, the non-zero unstable fixed point is 
$x^{(m) *}$
and unstable and $x^{(m) *}> \beta^{(m)}$.
Here, we define $a_0 \equiv 1,
 \alpha ^{(0)}\equiv \alpha,
\beta ^{(0)}\equiv \beta$.\par

{\bf Proof}

First of all, we notice that 
$A_{\vecphi ^m (\veca)}(x ^{(m)})$ is really
a trapezoid map for $m \ge 1$
because from lemma 3'  $b<b^{(m)}<1$ for $a_m<a$
and $m\ge 1$. 

For the case $m=0$, the statement follows from 
lemma 1'.\par

Let us assume that the statement holds for $m\ge 0$.
Since $x^{(m) *}= \frac{\gamma ^{(m)} a^{(m)}}
{\gamma ^{(m)} a^{(m)}+1}$
is the unstable fixed point of
 $A_{\veca ^{(m)}} (x^{(m)})$ for $a_{m+1}<a$,
from lemma 2', taking $A_{\veca ^{(m)}}^2 (x^{(m)})$
and rescaling $x^{(m) }$ as 
$x^{(m+1)}=\frac{x^{(m) *} - x^{(m)}}{\Delta x^{(m)}}$,
we obtain $A_{\veca ^{(m+1)}} (x^{(m+1)})$
which is defined in $[0, 1],$
where $\Delta x^{(m)} =
 \gamma^{(m)} (a^{(m)}-1)/(1+ \gamma^{(m)} a^{(m)})$.
From the relation (2$\cdot$14) it follows that
$x^{(m) }_M = \beta ^{(m)}$ at $a=a_{m+1}$.
Let us consider $x ^{(m+1)}_M =
a^{(m+1)} \alpha^{(m+1)}$.
Since $x^{(m+1)}_M$
is the continuous increasing function w.r.t. $a$,
we obtain $\alpha^{(m+1)}<
x^{(m+1)}_M < \beta^{(m+1)}$
for $a_{m+1}<a<a_{m+2}$.  
This implies $x^{(m+1)}_M$ is the unique fixed point
for $A_{\vecphi ^{m+1} (\veca)}(x ^{(m+1)})$
and stable.
It is easily shown that 
for $a_{m+2}<a$,  $x^{(m+1)}_M$ is no more fixed point
but $x^{(m+1) *}= \frac{\gamma ^{(m+1)} a^{(m+1)}}
{\gamma ^{(m+1)} a^{(m+1)}+1}$
becomes unstable fixed point.  Further, 
it is easily shown that 
$x^{(m+1) *}> \beta^{(m+1)}$ for $a_{m+1}<a$, 
and $x^{(m+1)}_M=x^{(m+1) *}$  at $a=a_{m+2}$.
 Q.E.D. \\
\par
{\bf lemma 5'}

The unique non-zero fixed point of 
$A_{\vecphi ^m (\veca)}(x ^{(m)})$ in lemma 4'
is the periodic cycle with prime period $2^m$.\\

The proof is the same as that in
the symmetric case.
Thus, we obtain the following theorem.\\
\par
{\bf Theorem 2}

$x=0$ is period 1 solution of 
$A_{\veca}(x)$ for any $a>0$.  Except for this,
for any non-negative integer $m$, 
for $a_m<a<a_{m+1}$ there is 
the unique cycle with prime period  $2^k$ for $k=0,1,
\cdots,m$ and no other periodic 
points exist.  The periodic solution with
prime period $2^m$ is stable for
$a_m <a <a_{m+1}$ and unstable for
$a_{m+1}<a$.
At $a=a_{m+1}$, the period $2^m$ cycle and the
period $2^{m+1}$ cycle coincide.\\
\par

The uniqueness of the cycle with prime period  $2^k$ 
$k=0,1,2,
\cdots,m$ for $a_m<a<a_{m+1}$
is proved by the same argument as that
in the symmetric case.
Thus, likewise the symmetric case, 
we conclude that as $a$ is increased the 
period doubling bifurcation cascades,
and at $a=a_m$ the periodic solution with 
prime period $2^m$ appears.\par

Now, let us investigate the upper bound of the 
sequence $\{ a_m \}$.
For $m \ge 1$
 $x^{(m)}_M(a)$ is defined at least for $ a _1 < a$.
These are the strictly increasing
functions w.r.t $a$.  
For any $m \ge 1$, 
$x^{(m)}_M(a_m)=0$ and 
$\lim_{a \rightarrow \infty} x^{(m)}_M(a)=\infty$
hold. 
Therefore, for $m \ge 1$, we define $a_M(m)$ as the
unique solution of the equation 
\begin{equation}
 x^{(m)}_M(a)=1.
\end{equation}
For $m=0$, we define $a_M(0) =\frac{1}{\alpha}$
and $a_M \equiv a_M(0)$.
Then, we obtain the following lemma.\\
\par
{\bf lemma 6'}
\begin{eqnarray}
&& 1<a_1<a_2<\cdots<a_m<a_{m+1}<\cdots \\
&& <a_M(m+1)<a_M(m)\cdots<a_M (2)<a_M (1)<a_M \nonumber
\end{eqnarray}

{\bf Proof}

For $m \ge 1$, 
since $x_M ^{(m)}(a_m)=0,  a_m<a_M(m)$ follows.
For $m \ge 0$, 
$x_M ^{(m+1)}$ is  rewritten as
\begin{equation}
 x_M ^{(m+1)}=\gamma ^{(m)} \{ a^{(m)} \}^2 
\frac{1}{ a^{(m)}-1} ( x_M ^{(m)}-\beta^{(m)} ).
\end{equation}
Then, we obtain $x_M ^{(m+1)}(a_M(m))=\frac{ a^{(m)}}
{ a^{(m)}-1} >1$ for $m \ge 0$.
Therefore, $a_M(m+1)< a_M(m)$ for $m \ge 0$. 
Thus, we obtain 
$a_m<a_{m+1}<a_M(m+1)<a_M(m).$
Q.E.D. \\

From this, 
\begin{equation}
a_c \equiv \lim_{m \rightarrow 
\infty}a_m \le \lim_{m \rightarrow\infty}a_M(m)<a_M
\end{equation}
follows.  Therefore $a_c$ is finite.

Now, let us investigate the symbolic sequence.
The sequence of $R$ and $L$
for the period $2^m$ solution becomes
Metropolis-Stein-Stein(MSS) sequence.  
Let us prove this.

As in the case of the symmetric map,
for $m\ge 0$ we define the onset of the $2^n$-cycle 
for $A_{\vecphi ^m (\veca)}(x^{(m)})$
as $a^{(m)}_n$ and
the $i$-th orbit of the period $2^n$ cycle
 as $x^{(m)} _{n,i}$.
$a^{(0)}_n$ is equal to previously defined $a_n$.
For $m=0$, we often omit the superscript $(0)$.
As $x_{m,1}$ we set the largest $x$ value
among $2^m$ members of the periodic cycle.
Then, $x_{m,1}=x_M=a\alpha$ when the cycle is stable.
Further, the coordinate space $[0,1]$ of  $x^{(m)}$ 
into the three intervals as
$I^{(m)}_L \equiv [0, \alpha^{(m)}],
I^{(m)}_C \equiv (\alpha^{(m)}, \beta ^{(m)})$
and $I^{(m)}_R \equiv [\beta ^{(m)},1].$

Then, we obtain the corresponding lemma 7', 8'
and 9' for $A_{\veca}(x)$ to
the lemma 7, 8, and 9 for $T_{\veca}(x)$,
respectively.
We omit the   statements and proofs of these
lemmas, since they are almost the same as those for
$T_{\veca}(x)$.
Thus, we obtain the MSS sequence 
for the period $2^m$ solution.

Now, as in the symmetric case, 
we derive the equations for 
$a_m$ and $a_c$, 
and obtain the asymptotic expression for $\delta_m$.
Defining $s(x)$ as before,
$f_s(x)$ for $x\in I_L \cup I_R$ is defined as
\begin{equation}
f_s (x) = a( \eta + \nu x), 
\;\; \eta = \gamma s(x),\;\; \nu = 1 - (1+\gamma ) s(x).
\end{equation}
In the below, $s_i, \eta _i$ and $ \nu _i$
are values evaluated at $x=x_{m,i}$.
Starting from $x_{m,1}= \alpha a$, 
we obtain $x_{m,2^m}$ for $a_m \le a\le a_{m+1}$,
\begin{eqnarray}
x_{m,2^m} &=& f_{s_{2^m -1}} (x_{m,2^m -1})
 \equiv \sum_{l=0} ^{2^m -1} c_l ^{(m)} 
a ^{2^m -l}\equiv F_m (a),\\
c_l ^{(m)} &=& \eta_{l} \prod_{j=l+1}^{2^m -1}\nu _j
\;\; {\rm for }
\;\; 1 \le l \le 2^m -2, \nonumber \\
c_0 ^{(m)} &=& \alpha \prod_{j=1}^{2^m -1}
 \nu_j = \eta _0 \prod_{j=1}^{2^m -1} \nu_j,\nonumber \\
c_{2^m -1} ^{(m)} &=& \eta_{2^m-1},\nonumber
\end{eqnarray}
where $\eta _0 \equiv \alpha$.  Thus,
\begin{eqnarray}
 F_m(a_m) & = & \frac{ \alpha + \beta +(-1)^{m-1} b}{2},\\
 F_{m-1}(a_m) & = & \frac{ \alpha + \beta+(-1)^{m-1} b}{2}.
\end{eqnarray}
We define $\zeta _l$ and $ \hat{\zeta} _l$
as the numbers of 1 in $s_1, s_2,\cdots,s_{2^l -1}$
and in $s_1, s_2,\cdots,s_{2^l}$,
respectively\cite{Ue 98}.
Then, their expressions are
\begin{eqnarray}
\hat{\zeta} _l = (2^{l+1}+(-1)^l )/3 \;\; 
\mbox{\rm for}\; l \ge 0,\\
\zeta_{l}= \hat{\zeta}_l -\frac{1+(-1)^l}{2} (l \ge 0)
,\nonumber  \\
\zeta_l = \zeta_{l-1}+ \hat{\zeta} _{l-1} \;\; 
{\rm for}\; l \ge 1, \nonumber
\end{eqnarray}
where $\zeta_0 \equiv 0$.  
The following relations hold,
\begin{eqnarray}
&&  \sum  _{l=0}^{m-2} \hat{\zeta}_l
=\zeta_{m-1},\\
&& \zeta_{2n+1}=\frac{4^{n+1}-1}{3}=\hat{\zeta}_{2n+1}
(n\ge0)\;\; \mbox{: odd number},\nonumber \\
&& \zeta_{2n}=\frac{2(4^n-1)}{3}=\hat{\zeta}_{2n}-1
(n\ge 1)\;\; \mbox{: even number},\nonumber  \\
&&  \prod_{j=1}^{2^m -1} \nu _j=(- \gamma)^{\zeta _m}
\;\; \mbox{ for } m \ge 1. \nonumber
\end{eqnarray}
From these it follows that $\hat{\zeta}_{l}$
is  odd number for $l\ge 1$.
Then, from the last relation in eq.(3$\cdot$15),
we obtain for $l \ge 0$
 \[ r_l \equiv (- \gamma) ^{- \zeta_m} c_l ^{(m)}
=\{ \prod_{j=1}^{2^m -1}\nu_j \}^{-1}
\eta_{l} \prod_{j=l+1}^{2^m -1}\nu _j
=\eta _l \{ \prod_{j=1}^l \nu_j \}^{-1}.\]
That is, $r_l $ is 
$m$-independent as long as it is defined.
Thus, we get
\begin{eqnarray}
r_l &=& \eta_l \{ \prod_{j=1}^l \nu_j \}^{-1}
 \mbox{ for any } l (\ge 0),\\
 r_0 & = & \alpha, \;\;r_{2^m}= - s_{2^m}
 \gamma ^{1- \hat{\zeta} _m }\;(m \ge 0).
\end{eqnarray}
Thus, we define
\begin{equation}
G_m (a) \equiv (-\gamma)^{-\zeta_m} a^{-2^m} F_m(a)=
 \sum_{l=0} ^{2^m -1} r_l  a ^{-l}.
\end{equation}
Let us consider the condition for
the convergence of $G_m(a)$. 
Using the relations (3$\cdot$19) and 
\begin{equation}
r_l= -  \gamma ^{ \hat{\zeta} _m } r_{2^m +l} 
 \mbox{ for $1\le l \le 2^m -1\;\; ( m \ge 1)$}, 
\end{equation}
 we obtain the following
recursive relation for $m \ge 1$,
\begin{equation} 
G_{m+1}(a)=(1-\tilde{a}_m)G_m(a)
+\tilde{a}_m(\alpha - \gamma s_{2^m}),
\end{equation}
and then we obtain for $m \ge 1$
\begin{equation}
 G_m(a)=\alpha \prod_{l=0}^{m-1}
(1-\tilde{a_l})
+ \sum_{k=0}^{m-1} [\prod_{l=k+1}^{m-1}(1-\tilde{a_l}) ]
\tilde{a_k}(\alpha - \gamma  s_{2^k}),
\end{equation}
where $\tilde{a_l}=a ^{-2^l}
\gamma^{-\hat{\zeta_l}}=\frac{1}{\sqrt{a^{(l-1)}}}$.
Since $\tilde{a_l}=(a \gamma ^{2/3})^{-2^l}
\gamma^{-(-1)^l/3}$, putting $\hat{\gamma}=$ max
$(\gamma, \gamma ^{-1})$, we get
\begin{equation}
 \tilde{a_l} \le (a \gamma ^{2/3})^{-2^l}
\hat{\gamma}^{1/3}. 
\end{equation}
Thus, if $a\gamma^{2/3}>1, \lim_{m\rightarrow \infty}
G_m(a)$ converges.  That is, 
$G_{\infty}(z) \equiv \lim_{m \rightarrow \infty} G_m (z)$
is the analytic function for $|z|> \gamma^{-2/3}$.
\footnote{In the previous paper\cite{Ue 98}, as
the condition for the convergence of $G_{\infty}(z)$
we gave $|z|> \mbox{\rm max }(1,\gamma^{-1})$.
This is a sufficient condition.}
From eq.($3\cdot12$) we obtain
\begin{equation}
 G_m(a_m)= (-\gamma)^{-\zeta_m}  a_m^{-2^m}
\frac{ \alpha + \beta+(-1)^{m-1} b}{2},
\footnotemark
\end{equation}
\footnotetext{This is 
the equation ($3\cdot4$) in ref.\cite{Ue 98}.
 There are several misprints in \cite{Ue 98}.
$\zeta_{m}$ in the definition of $G_m(a)$ and 
eq.($3\cdot4$) should be changed to $-\zeta_{m}$.}
and then
\begin{equation}
G_{\infty} (a_c) = 0.
\footnotemark
\end{equation}
\footnotetext{Note that $a_c\gamma^{2/3}>1$ follows
from $a_1 \gamma >1$. Since
$a_c ^{- 2^m}\gamma ^{ -\hat{\zeta} _m }
= \tilde{a}_m (a=a_c)=(a_c \gamma ^{2/3})^{-2^m}
\gamma^{-(-1)^m/3}$ and
$a_c ^{- 2^m}\gamma ^{ -\zeta _m }
=a_c ^{- 2^m}\gamma ^{ -\hat{\zeta} _m }
\gamma ^{\frac{1+(-1)^m}{2}}$ these tend
to 0 as $m \rightarrow \infty$.}
Similar to the symmetric case,
we obtain from eq.(3$\cdot$23) 
\begin{eqnarray}
  \epsilon _m &=& \{ G_m (a_c)- \Gamma a_c ^{-2^m} \}
(1+ \bar{h}_m)/G'_{\infty} (a_c),\\
&& \Gamma =(-1)^{ -\zeta _m }
\frac{ \alpha + \beta+(-1)^{m-1} b}{2},\;\;
\bar{h}_m=\frac{G'_{\infty} (a_c)}
{G_m'(\bar{a}_m)+
 \Gamma 2^m (\hat{a}_m)^{-2^m -1}}-1, \nonumber
\end{eqnarray}
where $a_m <\bar{a}_m<a_c,\;\;a_m <\hat{a}_m<a_c$
 and $\lim_{m \rightarrow \infty}\bar{ h}_m =0$.
Further, using the relation (3$\cdot$17)  we get
\begin{eqnarray}
 G_m(a_c)& =& -  a_c ^{- 2^m}\gamma ^{ -\hat{\zeta} _m } 
 (\alpha - \gamma s_{2^m}) + 
(a_c ^{- 2^m}\gamma ^{ -\hat{\zeta} _m })^2 q_m,\\
&& q_m =-  \frac{1}{1-a_c ^{- 2^m}
\gamma ^{ -\hat{\zeta} _m }}(
 \alpha - \gamma s_{2^m} + \gamma  ^{ 2 \hat{\zeta} _m }
\sum_{l=0}^{\infty}r_{2^{m+1} + l}a_c ^{ - l}).\nonumber
\end{eqnarray}
Substituting (3$\cdot$26)  into (3$\cdot$25) , we obtain 
\begin{eqnarray}
  \epsilon _m &= & a_c - a_m 
=\frac{ b a_c ^{-2^m} \gamma ^{- \zeta _m }}
{G'_{\infty} (a_c)}(1 + h_m),\\
&& h_m= \bar{h}_m (1+ a_c ^{- 2^m}
 \gamma ^{ \zeta _m - 2 \hat{\zeta} _m } \frac{q_m}{b})
+ a_c ^{- 2^m}
 \gamma ^{ \zeta _m - 2 \hat{\zeta} _m } \frac{q_m}{b}.
\end{eqnarray}
$\lim_{m \rightarrow \infty} h_m=0$ is shown in appendix B.
Thus,
\begin{eqnarray}
 \delta_m & =&  \gamma ^{(-1)^m/3 } 
(a_c \gamma ^{2/3}) ^{2^m} (1+l_m),\\
&& l_m=[ h_m -h_{m+1}-a_c ^{- 2^m}
\gamma ^{ -\hat{\zeta} _m }(1+h_{m+1})
+ a_c ^{- 2^{m+1}} \gamma ^{ -\hat{\zeta} _{m+1} }
(1+h_{m+2})] \nonumber \\
&& /[1+h_{m+1} -  a_c ^{- 2^{m+1}} 
\gamma ^{ -\hat{\zeta} _{m+1} }(1+h_{m+2})],\nonumber
\end{eqnarray}
and $\lim_{m \rightarrow \infty} l_m=0$.

To obtain a similar relation to the relation (2$\cdot$32) 
in the case of symmetric map, we define 
$v_m(a,\gamma)$ as
\begin{equation}
v_m(a,\gamma)=1/u_m(a,\gamma)
=\prod_{l=0} ^{m-1}v(a^{(l)},\gamma ^{(l)}),
\;\; v(a,\gamma) = 
\frac{1}{u(a,\gamma)}=\frac{\gamma (a-1)}{\gamma a+1}.
\end{equation}
For $m \ge 1$, $v_m(a)$ is rewritten as
\begin{eqnarray}
v_m(a)&=&\frac{(1-a^{-1})}{g_{m-1}(a)}
\prod_{l=0} ^{m-2} h_l(a),\\
&& g_{l}(a)\equiv 1+ \gamma ^{-\hat{\zeta}_l} a^{-2^l},\;\;
 h_{l}(a)\equiv 1- \gamma ^{-\hat{\zeta}_l} a^{-2^l},
\mbox{ for $l \ge 0$} \nonumber,\\
&& \prod_{l=0} ^{-1} h_l(a) \equiv 1, \;\;
v_1 (a)=v(a).\nonumber
\end{eqnarray}
Then, for $m \ge 2$, 
the following equation is proved by using the
above relations
\begin{equation}
G_{m-1}(a) = \alpha +
\frac{1}{1+\gamma^{-1}}
\{ g_{m-1}(a) v_m(a)-1 -(-1)^m a^{-2^{m-1}}
\gamma ^{-\zeta_{m-1}} \}.
\end{equation}
See Appendix A.
Thus, the condition for $a_m$,
$v_m(a_m)=b$, is equivalent to the 
following  equation derived from eq.(3$\cdot$13), 
\begin{equation}
 G_{m-1}(a_m)=(-\gamma)^{-\zeta_{m-1}}a_m ^{-2^{m-1}}
\frac{\alpha +\beta +(-1)^{m-1}b}{2}.
\end{equation}

  From relation (3$\cdot$32),
we get for $a> \gamma^{-2/3}$
\begin{equation}
 G_{\infty}(a) =\alpha+ \frac{1}{1+\gamma^{-1}}
( v_{\infty}(a) -1).
\end{equation}
Thus,
\begin{eqnarray}
G_{\infty}'(a) &=&
 \frac{1}{1+\gamma^{-1}} v_{\infty}'(a)
= \frac{2b}{1+\gamma^{-1}} \tau(a),\\
&& \tau(a) \equiv 
\sum _{l=0} ^{\infty}
\frac{1+\gamma ^{(l)}}
{2(a^{(l)}-1)(\gamma ^{(l)}a^{(l)}+1)}
\gamma ^{2(2^l-(-1)^l)/3}
2^l a ^{2^l -1}>0.\nonumber
\end{eqnarray}
Therefore,  we obtain
\begin{equation} \epsilon_m =\frac{1}{2 \tau(a_c)}
 a_c  ^{-2^m}\gamma^{-\zeta_m}(1+\gamma^{-1})(1+h_m).
\end{equation}

\section{ Period doubling bifurcation of the
period $p$ solution : symmetric case }

In this section, we consider the period doubling
bifurcation of the periodic solution with
the prime period $p( \ge 3)$ which emerges by a
tangent bifurcation.

Let us consider the mapping $T_{\veca}^p(x)$.
At $a=a_M=\frac{2}{1-b}$, the map looks like Fig.5.

\begin{figure}[htb]
     \parbox{\halftext}{
     \epsfbox{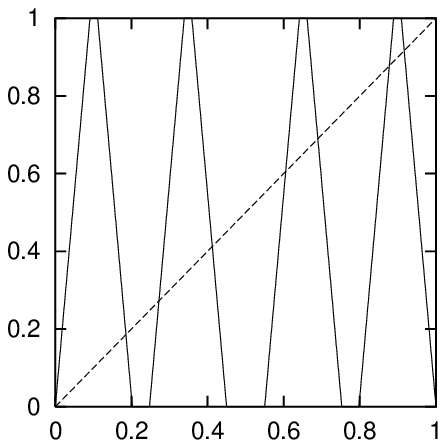}
 \caption{Trapezoid map $T^3 _{(a,b)}(x)$ for
$a=a_M$ with $b=0.1$.}}
 \label{fig:5}
\hspace{8mm}
     \parbox{\halftext}{
     \epsfbox{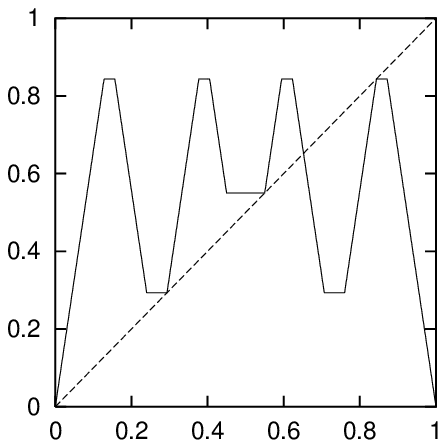}
 \caption{ Trapezoid map $T^3 _{(a,b)}(x)$ for
$a=a_{p,0}$ with $b=0.1$.}}
 \label{fig:6}
\end{figure}

As $a$ is decreased from $a=a_M$, at some
value of $a=a_{p,0}$, the map becomes
tangent to the line $y=x$, as is illustrated in Fig.6.
If we put $x_1=a\frac{1-b}{2}$, 
at this point the symbols for $x_1, x_2, \cdots, x_{p-1}$
are $RL\cdots L$.  Then, 
\begin{equation}
x_p=a^{p-1} ( 1- \frac{a(1-b)}{2}).
\end{equation}
As is shown in Appendix C, $x_p(a)$ takes
the maximum value at $a=a_{
 p,max }\equiv \frac{p-1}{p} a_M$
and  $\frac{1+b}{2} < x_p(a_{p,max})$ for $p \ge 3$.
Thus, the point $a_{p,0}$ where the tangent
bifurcation takes place satisfies the condition
\begin{equation}
x_p(a_{p,0})=\frac{1+b}{2}.
\end{equation}
This equation has two solutions, one is $a=1$ 
and the other corresponds to $a_{p,0}$.
Thus, we obtain
\[  1<  \frac{p-1}{p} a_M <a_{p,0}<a_{p,1},\]
where $a_{p,1}$ is defined by
\begin{equation}
x_p(a_{p,1})=\frac{1-b}{2},
\end{equation}
and $a_{p,1}>1$.  See Fig.7.

\begin{figure}
     \centerline{\epsfbox{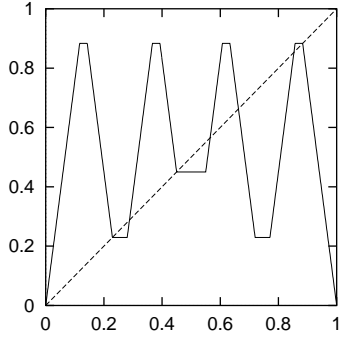}}
 \caption{Trapezoid map $T^3 _{(a,b)}(x)$ for
$a=a_{p,1}$ with $b=0.1$.}
 \label{fig:7}
\end{figure}

  $a_{p,1}$ is
the onset point of the period $2 p$ point.
The symbolic sequence 
for $x_1, x_2, \cdots, x_{p-1}$
is  $RL\cdots L$ for $a_{p,0}<a$. See Appendix D.
For $a_{p,0}< a <a_{p,1}$, let us consider
the map $T^p _{\veca}(x)$.
The unstable periodic point with  
period $p$ of $T_{\veca} (x)$
is 
\begin{equation}
x^*=\frac{a^{p-1}(a-1)}{a^p -1}.
\end{equation}
Rescaling the map $T^p _{\veca}(x)$  around
$x=1/2$ for $a>a_{p,0}$, we obtain the trapezoid map
$T_{\veca^{(0)}}(x^{(0)})$, where
\begin{eqnarray}
\veca^{(0)}&=&(a^{(0)}, b^{(0)}),\;\;
a^{(0)}=a^p,\;\; b^{(0)}=\frac{b}{\Delta x}
=r(a)b,\;\; \Delta x=2 x^* -1,\\
r(a)&=&\frac{1}{\Delta x}=
\frac{a^p-1}{a^p-2a^{p-1}+1},
\;\;x^{(0)}=\frac{x^*-x}{2 x^*-1}.\nonumber
\end{eqnarray}
For $a>1$, $r'(a)<0$.
As before, we define $x^{(0)}_M= a^{(0)}
 \frac{1-b^{(0)}}{2}$. 
For $a_{p,0}<a<a_{p,1}$, $x^{(0)}_M$
corresponds to $x_p$, that is
$x^{(0)}_M=\frac{x^*-x_p}{2 x^*-1}$.
Therefore, at $a=a_{p,0}, x^{(0)}_M=0$
and at $a=a_{p,1}, x^{(0)}_M=\frac{1+b^{(0)}}{2}$.
We note that $b^{(0)}(a_{p,0})=1.$
Since $r(a)$ is
the strictly decreasing continuous function
w.r.t. $a$ for $a>1$, 
$b^{(0)}(a)$ is
the strictly decreasing continuous function
and $x^{(0)}_M (a)$ is 
the strictly increasing continuous function
w.r.t. $a$.
Therefore, for $a_{p,0}<a<a_{p,1},
\frac{1-b^{(0)}}{2}<x^{(0)}_M <\frac{1+b^{(0)}}{2}$
, and for $ a_{p,1}<a, 
\frac{1+b^{(0)}}{2}<x^{(0)}_M $.
That is, $x^{(0)}_M $ is the stable fixed point of
$T_{\veca^{(0)}}(x^{(0)})$ for  $a_{p,0}<a<a_{p,1}$,
and for $ a_{p,1}<a$ the
fixed point is $x_u ^* \equiv \frac{a^{(0)}}{a^{(0)}+1}$
and is unstable.
For $ a_{p,1}<a$, rescaling the map 
$T_{\veca^{(0)}} ^2 (x^{(0)})$
we obtain the trapezoid map  $T_{\veca^{(1)}}(x^{(1)})$,
where,
\begin{eqnarray}
\veca ^{(1)}& \equiv & \vecphi (\veca ^{(0)})
=(a^{(1)},b^{(1)}),\;\;
a^{(1)}=\{a^{(0)}\} ^2 = a^{2p},\\
b^{(1)}&=&u(a^{(0)})b^{(0)}=u(a^p)r(a)b,\;\;
u(a)=\frac{a+1}{a-1}.\nonumber
\end{eqnarray}
As before, we define $\veca ^{(m)}$ as
\begin{eqnarray}
\veca ^{(m)}& \equiv & \vecphi^m (\veca ^{(0)})
=(a^{(m)},b^{(m)}),\;\;
a^{(m)}=\{a^{(m-1)}\} ^2 = a^{2^m p},\\
b^{(m)}&=&u(a^{(m-1)})b^{(m-1)}=u_m(a)r(a)b,\;\;
u_m(a)=\prod_{l=0}^{m-1}u(a^{(l)}).\nonumber
\end{eqnarray}
Since $b^{(m)}(a)$ is the decreasing function w.r.t.
$a$, we can repeat the same argument as in section 2.
At $a=a_{p,m}$, the period $2^m$ solution
of $T_{\veca^{(0)}} (x^{(0)})$ appears and
$a_{p,m}$ satisfies the condition
\begin{equation}
 b^{(m)}(a_{p,m})=1,\;\;\mbox{\rm for } m \ge 0,
\end{equation}
and 
\begin{equation}
1<a_{p,0}<a_{p,1}<a_{p,2}<\cdots <\frac{2}{1-b}=a_M.
\end{equation}
The symbolic sequence for the period $2^m$ solution
is the MSS sequence for the map 
 $T_{\veca^{(0)}} (x^{(0)})$.
For $a_{p,m}\le a \le a_{p,m+1}$,
starting from $x^{(0)}_1=x^{(0)}_M$,
$x^{(0)}_{2^m}$ is expressed by the function
$F_m(a^{(0)}, b^{(0)})$ .
Here, $F_m(a, b)$ is defined by eq.($2\cdot10$)
and
we include $b$ as an independent variable to $F_m$
explicitly.
Thus, we obtain the other condition for
$a_{p,m}$,
\begin{equation}
 F_m(a^{(0)}(a_{p,m}),b^{(0)}(a_{p,m}))
  =  \frac{ 1+(-1)^{m-1} b^{(0)}(a_{p,m})}{2}.
\end{equation}
In terms of $ G_m(a,b)$ defined by eq.(2$\cdot$19),
the equation (2$\cdot$21)  becomes
\begin{equation}
 G_m(a^{(0)}(a_{p,m}),b^{(0)}(a_{p,m}))=
a^{(0)}(a_{p,m}) ^{-2^m} 
\frac{ (-1)^m -b^{(0)}(a_{p,m})}{2},
\end{equation}
and we obtain
\begin{equation}
G_{\infty} (a^{(0)}(a_{p,c}),b^{(0)}(a_{p,c}))=0,
\end{equation}
where $a_{p,c}= \lim_{m \rightarrow \infty}a_{p,m}$.
From these equations, we get for
$\epsilon _{p,m} = a_{p,c} - a_{p,m}$,
\begin{equation}
 \epsilon_{p,m} \simeq
\frac{ a_{p,c} ^{-p 2^m} r(a_{p,c}) b }
{ \frac{\partial G_{\infty}}{\partial a^{(0)}}
 p a_{p,c}^{p-1} + \frac{1}{2}|r'( a_{p,c})|b}.
\end{equation}
Since $\frac{\partial G_{\infty}(a,b)}{\partial a}>0$,
the denominator is positive.
Therefore, we obtain
\begin{equation} \delta _m \sim a_{p,c}^{p 2^m}.
\end{equation}
As for the relations among $ a_{p,0}$ s, 
the following ordering holds,
\begin{equation}
 a_c \le a_{3,0}<a_{4,0}<\cdots
 < a_{p,0}< a_{p+1,0}<\cdots <a_M.
\end{equation}
See  Appendix E.\\

\section{ Period doubling bifurcation of the
period $p \ge 3$ solution : asymmetric case }

As in the previous case, we investigate the
period doubling bifurcation of the period $p(\ge 3)$ 
solution in the asymmetric case.

In this case also,  we consider the period $p$
solution starting with $x_1=a \alpha$
with the sequence of symbols, $RL\cdots L$.
Then, we obtain
\begin{equation}
x_p=\gamma a^{p-1} ( 1- a \alpha).
\end{equation}
As is shown in Appendix C,
$x_p(a)$ takes
the maximum value at $a=a_{ p,max }
\equiv \frac{p-1}{p} a_M$
and $\beta < x_p(a_{p,max})$ for $p \ge 3$,
where $a_M= 1/\alpha$ as before.  
Then, the onset point $a_{p,0}(>a_{p,max})$ 
of the period $p$ solution
satisfies
\begin{equation}
x_p(a_{p,0})=\beta.
\end{equation}
Although this equation has two solutions, 
we should adopt the larger one as in the symmetric case.
The symbolic sequence for $x_1, x_2, \cdots, x_{p-1}$
is  $RL\cdots L$ for $a_{p,0}\le a$. See Appendix D.
On the other hand, the onset point of the
period $2p$ solution,  $a_{p,1}$, satisfies
\begin{equation}
x_p(a_{p,1})=\alpha.
\end{equation}
Then, for $a_{p,0}< a $,
the unstable periodic point $x^*$ 
with period $p$ for $A _{\veca}(x)$
is
\begin{equation}
x^*=\frac{(\gamma a -1) \gamma a^{p-1}}
{\gamma ^2 a^p -1}.
\end{equation}
Similar to the symmetric case,
we obtain an asymmetric trapezoid map
$A_{\veca^{(0)}}(x^{(0)})$ by
rescaling the map $A^p _{\veca}(x)$
in the vicinity of $I_C$ for $a>a_{p,0} $, where
\begin{eqnarray}
\veca^{(0)}&=&(a^{(0)}, b^{(0)},\gamma ^{(0)}),\;\;
a^{(0)}=\gamma ^2a^p,\;\;
\gamma^{(0)}=\gamma ^{-1},\;\;
 b^{(0)}=\frac{b}{\Delta x}
=r(a)b,\\
 \Delta x&=&\frac{\gamma  g(a)}{\gamma ^2 a^p -1},\;\;
g(a)=\gamma a^p-(1+\gamma)a^{p-1}+1,\nonumber \\
r(a)&=&1/\Delta x = \frac{\gamma ^2 a^p -1}{\gamma g(a)},
\;\; x^{(0)}=\frac{x^*-x}{\Delta x},\nonumber \\
\alpha ^{(0)}&\equiv & \frac{x^* -\beta}{\Delta x}
=\frac{\gamma ^2a^p(1-\beta) - \gamma a^{p-1}+\beta}
{\gamma g(a)},\nonumber \\
\beta ^{(0)}&\equiv& \frac{x^* -\alpha}{\Delta x}
=\frac{\gamma ^2a^p(1-\alpha) - \gamma a^{p-1}+\alpha}
{\gamma g(a)}.\nonumber
\end{eqnarray}
$r'(a)$ becomes
\begin{equation}
r'(a)=- \frac{(1+ \gamma) a^{p-2} f(a)}{\gamma g(a)^2},
\end{equation}
where $f(a)= \gamma ^2 a^p -p \gamma a +p-1$.
We can prove that at least for $a_{p,0} \le a$,
$f(a) >0$.  Thus, for $a_{p,0} \le a$, $r'(a)<0$.
Further, $\alpha ^{(0) '} (a)=\frac{b a^{p-2}f(a)}{\gamma
g(a)^2} >0$ and $\beta ^{(0)'}(a)=-\gamma \alpha ^{(0)'}(a)
<0$ for $a\ge a_{p,0}$.  
 Thus, $b^{(0)}$ is
the strictly decreasing continuous function and
$x_M ^{(0)}=a^{(0)} \alpha ^{(0)}$ 
is the strictly increasing continuous
function w.r.t. $a$.  At $a=a_{p,0}$,
$\alpha ^{(0)}=0, \beta ^{(0)}= b^{(0)}=1$.
For $a_{p,0} \le a \le a_{p,1}$
$x_p(a)$ corresponds to $x_M ^{(0)}$,
that is, $x_M ^{(0)}= \frac{x^*-x_p}{\Delta x}$.
Then, at $a = a_{p,0}, x_M ^{(0)}=0$
and at $a = a_{p,1}, x_M ^{(0)}=\beta  ^{(0)}$.
Thus, $\alpha ^{(0)}<x_M ^{(0)}<\beta ^{(0)}$
for $a_{p,0}<a<a_{p,1}$,
and $x_M ^{(0)}$ is the stable fixed point
for $A_{\veca^{(0)}}(x^{(0)})$.
Further, for $a_{p,1}<a$,  $x_M ^{(0)}>\beta ^{(0)}$
and the fixed point becomes
$x_u ^* \equiv \frac{\gamma^{(0)}a^{(0)}}
{1+\gamma^{(0)}a^{(0)}}$ and unstable.
For $a_{p,1}<a$, rescaling the map 
$A_{\veca^{(0)}}^2 (x^{(0)}) $, 
we obtain $A_{\veca^{(1)}}(x^{(1)})$, where
\begin{eqnarray}
\veca ^{(1)}& \equiv & \vecphi (\veca ^{(0)})
=(a^{(1)},b^{(1)}, \gamma^{(1)}),\;\;
a^{(1)}=\{\gamma ^{(0)} a^{(0)}\} ^2 = \gamma ^2a^{2p},\\
b^{(1)}&=&u(a^{(0)}, \gamma ^{(0)})b^{(0)}
=u(a^p)r(a)b,\;\; \gamma^{(1)}=\{ \gamma^{(0)}\}^{-1},\;\;
u(a,\gamma)=\frac{\gamma a+1}{\gamma(a-1)}.\nonumber
\end{eqnarray}
Thus, $a^{(1)}$ is increasing, $b^{(1)}$ is
decreasing and $\gamma^{(1)}$ is constant w.r.t. $a$.
Further, we define  $\veca ^{(m)}$ as
\begin{eqnarray}
\veca ^{(m)}& \equiv & \vecphi^m (\veca ^{(0)})
=(a^{(m)},b^{(m)}, \gamma ^{(m)}),\;\;
a^{(m)}=\{\gamma ^{(m-1)} a^{(m-1)}\} ^2 ,\\
b^{(m)}&=&u(a^{(m-1)},\gamma ^{(m-1)})b^{(m-1)}=
u_m(a, \gamma )r(a)b,\;\;
u_m(a, \gamma)=\prod_{l=0}^{m-1}u(a^{(l)},\gamma ^{(l)}),
\nonumber \\
 \gamma^{(m)}&=& \{ \gamma ^{(m-1)}\}^{-1}.\nonumber
\end{eqnarray}

Since $b^{(m)}$ is the decreasing function
w.r.t. $a$, we can make a quite similar argument to that
in \S4.
In particular, at $a=a_{p,m}$, the period doubling
bifurcation takes place and 
\begin{equation} 
b^{(m)}(a_{p,m})=1\;\;\mbox{\rm for } m \ge 0,
\end{equation}
and
\begin{equation} 
1<a_{p,0}<a_{p,1}<a_{p,2}<\cdots <\frac{1}{\alpha}=a_M.
\end{equation}
The symbolic sequence for the period $2^m$ solution
is the MSS sequence for the map $A_{\veca^{(0)}}(x^{(0)})$.
For $a_{p,m}\le a \le a_{p,m+1}$,
starting from $x^{(0)}_1=x^{(0)}_M$,
$x^{(0)}_{2^m}$ is expressed by the function
$F_m(a^{(0)}, b^{(0)}, \gamma^{(0)})$  and we get
\begin{equation}
 F_m(a^{(0)}(a_{p,m}), b^{(0)}(a_{p,m}), \gamma ^{(0)})
  =  \frac{ \alpha ^{(0)}(a_{p,m}) + \beta ^{(0)}(a_{p,m}) 
+(-1)^{m-1} b^{(0)}(a_{p,m})}{2},
\end{equation}
where $ F_m(a,b,\gamma)$ is defined by eq.(3$\cdot$11). 
In terms of $G_m(a,b,\gamma)$ defined
by eq.(3$\cdot$18), the equation (5$\cdot$11) becomes
\begin{eqnarray}
&& G_m(a^{(0)}(a_{p,m}),b^{(0)}(a_{p,m}), \gamma ^{(0)})
= \\
&& (-\gamma^{(0)}) ^{-\zeta_m} \{a^{(0)}_m
(a_{p,m}) \} ^{-2^m}
\frac{ \alpha^{(0)}(a_{p,m})
 + \beta^{(0)}(a_{p,m}) +(-1)^{m-1} 
b^{(0)}(a_{p,m})}{2}. \nonumber
\end{eqnarray}
Let us define
$a_{p,c}= \lim_{m \rightarrow \infty}a_{p,m}$.
Thus, we obtain
\begin{equation}
G_{\infty} (a^{(0)}(a_{p,c}),b^{(0)}(a_{p,c})
,\gamma ^{(0)} )=0,
\end{equation}
and $\epsilon_{p,m} = a_{p,c} - a_{p,m}$ is
given by
\begin{equation}
 \epsilon_{p,m} \simeq
\frac{(\gamma ^2 a_{p,c}^p) ^{-2^m}
 \gamma ^{ \zeta _m }  r(a_{p,c}) b }
{ \frac{\partial G_{\infty}}{\partial a^{(0)}}
\gamma ^2 p a_{p,c}^{p-1} +
\frac{1}{2}|r'( a_{p,c})|b},
\end{equation}
and then
\begin{equation}
 \delta_m \simeq  \gamma ^{(-1)^{m-1}/3 }
 (a_{p,c}^p \gamma ^{4/3}) ^{2^m}.
\end{equation}
As for the relations among $ a_{p,0}$ s,
the following ordering holds,
\begin{equation}
 a_c \le a_{3,0}<a_{4,0}<\cdots
 < a_{p,0}< a_{p+1,0}<\cdots <a_M.
\end{equation}
See  Appendix E.\\

\section{ Summary and discussion}

In this paper, we studied the symmetric
and the asymmetric trapezoid maps rigorously.
We gave the proofs of several results
about the period doubling bifurcation which
occurs as the parameter $a$ is increased.
We obtained following scaling results for
the period doubling
bifurcation starting from a period one solution.
\begin{enumerate}
\item Symmetric case 
\begin{eqnarray*}
\epsilon _m &=& \frac{b a_c ^{-2^m}}{G'_{\infty} (a_c)}
(1+h_m),\\
\delta_m &=& a_c ^{2^m}(1+l_m)
\end{eqnarray*}
where $\lim_{m \rightarrow \infty} h_m=0$
and $\lim_{m \rightarrow \infty} l_m=0$.
\item Asymmetric case
\begin{eqnarray*}
  \epsilon _m &=&
\frac{ b a_c ^{-2^m} \gamma ^{- \zeta _m }}
{G'_{\infty} (a_c)}(1 + h_m),\\
\delta_m & =& \gamma ^{(-1)^m/3 }
(a_c \gamma ^{2/3}) ^{2^m} (1+l_m),
\end{eqnarray*}
where $\lim_{m \rightarrow \infty} h_m=0$
and $\lim_{m \rightarrow \infty} l_m=0$.
\end{enumerate}
The new results in this paper are on the period
doubling bifurcation which starts from
the periodic solution with any prime period $p (\ge 3)$.
\begin{enumerate}
\item Symmetric case 
\begin{eqnarray*}
\epsilon_{p,m} & \simeq &
\frac{ a_{p,c} ^{-p 2^m} r(a_{p,c}) b }
{ \frac{\partial G_{\infty}}{\partial a^{(0)}}
 p a_{p,c}^{p-1} + \frac{1}{2}|r'( a_{p,c})|b}.\\
\delta_m &\simeq & a_{p,c}^{p 2^m}.
\end{eqnarray*}
\item Asymmetric case
\begin{eqnarray*}
\epsilon _m & \simeq & 
\frac{(\gamma ^2 a_{p,c}^p) ^{-2^m}
 \gamma ^{ \zeta _m }  r(a_{p,c}) b }
{ \frac{\partial G_{\infty}}{\partial a^{(0)}}
\gamma ^2 p a_{p,c}^{p-1} +
\frac{1}{2}|r'( a_{p,c})|b},\\
 \delta_m &\simeq & 
\gamma ^{(-1)^{m-1}/3 }
 (a_{p,c}^p \gamma ^{4/3}) ^{2^m}.
\end{eqnarray*}
\end{enumerate}
These results imply that
 for any period doubling cascade,
the accumulation rate to the 
accumulation point is extremely fast, in fact,
it is exponential.

In a one-dimensional
map with one hump, the Feigenbaum constant
depends on the power $z$
which characterizes the behaviour of the map
in the vicinity of the critical point.
It has been proved that $\lim _{z \rightarrow \infty}
\delta(z)= $ finite\cite{Eckmann 90}.
The result in this paper shows that 
$\delta(\infty) $ is infinity.
That is, the superconvergence takes place 
only in the case $z=\infty$.  \\
This feature is considered to be attributed 
to the flatness of the summit.
For example, in a one humped map 
with a flat summit,
superconvergence will take place if 
it satisfies some conditions, e.g.  the
absolute value of the derivative 
is greater than some constant
$\lambda >1$ outside a region 
which contains the flat part of the map.

In the other papers 
\cite{Beyer 82,Wang 87,Wang and Beyer 98}, 
similar results were obtained by a different method.
In those studies, the authors estimated 
quantities in question by inequalities and
obtained that $a_m$ is quadratically convergent,
in the case of period doubling of period one solution.
This implies $\lim_{m \rightarrow \infty}
\frac{ \ln \delta_m}{ \ln \delta_{m-1}}=2.$
  On the other hand,
our method is constructive.
In fact, we gave the precise equation for
the onset point $a_m$ of the periodic point
with period $2^m$ and also gave the
equation for the accumulation point $a_c$.
From these equations, we obtained the above 
scaling relations.
We also extended the argument not only to the asymmetric 
trapezoid map but also 
period doubling of the prime period $p( \ge3)$
solution which emerges from a tangent bifurcation.

In \cite{Wang 87},  the
authors mention the problem of convergence of
the sequence 
$\Delta \epsilon_m /(\Delta \epsilon_{m-1})^2$, 
and from our result, this is
$\epsilon_m /\epsilon_{m-1}^2 \simeq G_{\infty}'(a_c)
\gamma^{(-1+(-1)^m)/2} /b$ and does not
converge for the asymmetric case.

The trapezoid maps studied here are a kind of exactly 
solvable models of the renomalization group.
It is interesting to investigate the further
detailed bifurcation structures in these models.
This is a future problem.

\appendix
\section{Proof of relations ($2\cdot32$) and
 ($3\cdot32$)}

In this appendix, 
we prove the relation (3$\cdot$32). 
The relation (2$\cdot$32) is
obtained by putting $\gamma =1$.

For $m \ge 2$, the relation ($3\cdot32$) is
\begin{eqnarray}
 G_{m-1}(a)& = & \alpha +
\frac{1}{1+\gamma^{-1}}
\{ g_{m-1}(a) v_m(a)-1 -(-1)^m a^{-2^{m-1}}
\gamma ^{-\zeta_{m-1}} \},\\
&& v_m(a)=\frac{(1-a^{-1})}{g_{m-1}(a)}
\prod_{l=0} ^{m-2} h_l(a),\nonumber\\
&& g_{l}(a)\equiv 1+ \gamma ^{-\hat{\zeta}_l} a^{-2^l},\;\;
 h_{l}(a)\equiv 1- \gamma ^{-\hat{\zeta}_l} a^{-2^l},
\mbox{ for $l \ge 0$} \nonumber.
\end{eqnarray}

Let us define $\phi_m$ for $m \ge 1$ as
\begin{equation}
\phi_m(a)\equiv g_m(a) v_{m+1}(a)=
(1-a^{-1}) \prod_{l=0} ^{m-1} h_l(a).
\end{equation}
We expand $\phi_m(a)$ as
\begin{equation}
\phi_m(a)\equiv \sum_{l=0}^{2^m}w_l  ^{(m)}a^{-l},
\end{equation}
and then using relations (3$\cdot$15) we obtain for $m \ge 1$,
\begin{equation}
w^{(m)}_0=1,\;\; w^{(m)}_{2^m}=(-1)^{m-1}
\gamma ^{- \zeta _m}.
\end{equation}
In particular,
\begin{equation}
w^{(1)}_1=-(1+\gamma ^{-1}).
\end{equation}
Let us derive the recursive relations for $w^{(m)}_l$.
For $m \ge 1$,
\begin{equation}
\phi_{m+1}(a)=g_m(a) v_{m+1}(a) =h_m(a) \phi_m(a)
=\phi_m(a) - \gamma ^{- \hat{\zeta} _m} a^{-2^m}\phi_m(a).
\end{equation}
Comparing coefficients, we obtain for $m \ge 1$,
\begin{eqnarray}
&&{\cal O} (a^{-2^{m+1}}): \;\;  w^{(m+1)}_{2^{m+1}}
= - \gamma ^{ -\hat{\zeta}_m } w^{(m)}_{2^m},\\
&&{\cal O} (a^{-2^m}): \;\;  w^{(m+1)}_{2^m}
= - w^{(m)}_{2^m}-\gamma ^{ -\hat{\zeta}_m } w^{(m)}_0,\\
&&{\cal O} (a^{-(2^m+l)}): \;\;  w^{(m+1)}_{2^m+l}
=  -\gamma ^{ -\hat{\zeta}_m } w^{(m)}_l
\; \mbox{ for } 0 < l < 2^m,\\
&&{\cal O} (a^{-l}): \;\; w^{(m+1)}_l= w^{(m)}_l
\; \mbox{ for } 0 \le l < 2^m.
\end{eqnarray}
Using the relation (A$\cdot$4), eq.(A$\cdot$7) 
is automatically satisfied.
Form eq.(A$\cdot$8), we get
\begin{equation}
 w^{(m+1)}_{2^m}
= - ( 1 + \gamma ^{-1}) s_{2^m} 
\gamma ^{ 1 -\hat{\zeta}_m }.
\end{equation} 
Now, let us return to the quantity $G_{m-1}(a)$.
We put the r.h.s. of relation (A$\cdot$1)  
as $\hat{G}_{m-1}(a)$,
and expand it by $a^{-1}$ for $ m \ge 2$.
\begin{equation}
\hat{G}_{m-1}(a)=  \sum_{l=0}^{2^{m-1} -1} 
\bar{r}_l ^{(m-1)} a^{-l}.
\end{equation} 
Thus, we obtain the following relations for $ m \ge 2$,
\begin{eqnarray}
\bar{r}_0 ^{(m-1)}&=&\alpha,\\
\bar{r}_l ^{(m-1)}&=& \frac{w_l ^{(m-1)}}{1+ \gamma ^{-1}},
\;\; \mbox{ for } 0< l \le 2^{m-1} -1. 
\end{eqnarray}
Thus, from (A$\cdot$11), (A$\cdot$9) and (A$\cdot$10)
 we obtain for $m \ge 1$
\begin{eqnarray}
\bar{r}_{2^m}^{(m+1)}
&=& - s_{2^m} \gamma ^{1- \hat{\zeta}_m},\\
\bar{r}_{l+2^m} ^{(m+1)},
 &=& - \gamma ^{- \hat{\zeta}_m}\bar{r}_l ^{(m)}
\;\; \mbox{ for } 0< l \le 2^m -1,\\
\bar{r}_l^{(m+1)} &=& \bar{r}_l^{(m)} \;\; 
\mbox{ for } 0 \le l \le 2^m -1.
\end{eqnarray}
The relation  (A$\cdot$17) implies  that
$\bar{r}_l ^{(m)}$ is $m$ independent
as long as it is defined and so we omit the superscript
$(m)$.  
From relations (A$\cdot$15) and (A$\cdot$16) it turns out
that  $\bar{r}_{l}$ for $l =2^m +1, \cdots, 2^{m+1} -1$
are determined by $\bar{r}_{l}$ 
 for $l =1, \cdots, 2^m-1$.
Thus, $ \bar{r}_{1}$ and $\bar{r}_{2^m}$ 
for $ (m \ge 1)$
determines $\bar{r}_{l}$ for $l>0$.
From eqs.(A$\cdot$5) and (A$\cdot$14) 
$\bar{r}_{1}=-1=r_1$ follows.
Also, from eqs.(3$\cdot$17) and (3$\cdot$19) we note
$r_l$ satisfies the same relations as (A$\cdot$15) 
and (A$\cdot$16).
Then, $\bar{r}_{2^m}= r_{2^m}$ for $ (m \ge 1)$.
  Therefore,
 $r_{l}=\bar{r}_{l}$ follows for $l \ge 1$.
Finally, from the relation (A$\cdot$13) 
we have $\bar{r}_{0}= \alpha =r_0$.
Therefore, we obtain $\hat{G}_{m-1}(a)=G_{m-1}(a)$.
That is, the relation (3$\cdot$32) holds.\\

\section{Proof of $\lim_{m \rightarrow \infty} h_m=0$}

In this appendix, we prove
\[ \lim_{m \rightarrow \infty} h_m=0,\]
where $h_m$ is
\begin{eqnarray*}
 h_m & = & \bar{h}_m (1+ a_c ^{- 2^m}
 \gamma ^{ \zeta _m - 2 \hat{\zeta} _m } \frac{q_m}{b})
+ a_c ^{- 2^m}
 \gamma ^{ \zeta _m - 2 \hat{\zeta} _m } \frac{q_m}{b},\\
&& q_m =-  \frac{1}{1-a_c ^{- 2^m}
\gamma ^{ -\hat{\zeta} _m }}(
 \alpha - \gamma s_{2^m} + \gamma  ^{ 2 \hat{\zeta} _m }
\sum_{l=0}^{\infty}r_{2^{m+1} + l}a_c ^{ - l}).
\end{eqnarray*}
To show this, we only have to prove that
\[  \lim_{m \rightarrow \infty} a_c ^{- 2^m} 
\gamma ^{ \zeta _m - 2 \hat{\zeta} _m } q_m =0.\]
First we estimate the series $S \equiv 
\sum_{l=0}^{\infty}r_{2^{m+1} + l}a_c ^{ - l}$.\\
Let $\tau_l$ be the number of 1 in $s_1, s_2,\cdots, s_l$.
  Then, $r_l$ is
\begin{equation}
r_l= \gamma s_l [ \prod_{j=1} ^l \{ 1-(1+\gamma)s_j \}
 ]^{-1} = \gamma s_l(-\gamma)^{-\tau _l}.
\end{equation}
Thus,
\[ | r_l| \le \gamma ^{1-\tau _l}.\]
Let $\tilde{\tau}_l$ 
be the number of 1 in $s_{2^{m+1}+1}, s_{2^{m+1}+2},
\cdots, s_{2^{m+1}+l}$.  Then,  for $l\ge 0$ we obtain
\[ \tau_{2^{m+1}+l}=\hat{\zeta}_{m+1}+\tilde{\tau}_{l},
\;\; \tilde{\tau}_{l}\le l,\;\;\tilde{\tau}_{0}\equiv 0.\]
Then,
\[ |S| \le \sum_{l=0}^{\infty}
\gamma ^{1- \hat{\zeta}_{m+1}
- \tilde{\tau}_{l}}a_c ^{ - l}. \]
Since $a_c \gamma >1$, we obtain
\begin{eqnarray*}
|S| & \le& \frac{a_c}{a_c -1} \gamma
^{1- \hat{\zeta}_{m+1}},\;\; \mbox{ for } \gamma \ge 1,\\
|S|  & \le & \frac{a_c \gamma }{a_c \gamma -1} \gamma
^{1- \hat{\zeta}_{m+1}},\;\; \mbox{ for } \gamma < 1.
\end{eqnarray*}
Thus for any $\gamma$,
\[ |S|  \le {\rm const. } \gamma^{- \hat{\zeta}_{m+1}}.\]
From the following relation
\[\zeta_m - \hat{\zeta}_{m+1} = - \hat{\zeta}_m 
- \frac{1+(-1)^{m-1}}{2},\]
we get
\[  a_c ^{- 2^m} \gamma ^{ \zeta _m} |S| 
\le {\rm const.} a_c ^{- 2^m}\gamma ^{ -\hat{\zeta} _m }
\gamma ^{- \frac{1+(-1)^{m-1}}{2}}.\]
Further, 
\[\zeta_m - 2\hat{\zeta}_m = - \hat{\zeta}_m 
- \frac{1+(-1)^m}{2},\]
and then
\[ a_c ^{- 2^m}\gamma ^{\zeta_m -2\hat{\zeta} _m }
= a_c ^{- 2^m}\gamma ^{- \hat{\zeta} _m}
\gamma^{- \frac{1+(-1)^m}{2}}.\]
Thus,  since $a_c ^{- 2^m}\gamma ^{- \hat{\zeta} _m}$
tends to 0 as $m \rightarrow \infty$,
we obtain
\[  \lim_{m \rightarrow \infty} a_c ^{- 2^m} 
\gamma ^{ \zeta _m - 2 \hat{\zeta} _m } q_m =0.\]

\section{Proofs of $a_{ p,max }= \frac{p-1}{p} a_M$ 
and $ x_p(a_{p,max}) > \beta $ for $p \ge 3$}

In this appendix, we prove that for the asymmetric
case, $x_p=\gamma a^{p-1} ( 1- a \alpha)$ has
the maximum value at $a=a_{ p,max }
\equiv \frac{p-1}{p} a_M$ and 
$ x_p(a_{p,max}) > \beta $ for $p \ge 3$.
Similar results for the symmetric case 
is obtained by putting $\gamma =1$.

Differentiating $ x_p(a)$ w.r.t. $a$, we get
\begin{equation}
 x_p'(a)= \gamma a^{p-2}(p-1)(1- \frac{p}{p-1} a \alpha).
\end{equation}
Then, at $a=a_{ p,max }\equiv \frac{p-1}{p} a_M$,
 $ x_p(a)$ has the maximum value 
\begin{equation}
 x_p(a_{ p,max })= \frac{\gamma}{p}
(\frac{p-1}{p})^{p-1}a_M ^{p-1}
=\frac{1}{p(1-\beta)}
(\frac{p-1}{p})^{p-1}\alpha ^{2-p}.
\end{equation}
Since $0<\alpha<\beta<1$ and $p\ge 3$,
\begin{equation}
 x_p(a_{ p,max })>\frac{1}{p(1-\beta)}
(\frac{p-1}{p})^{p-1}\beta ^{2-p}
=\frac{1}{p}
(\frac{p-1}{p})^{p-1}\frac{\beta}{g(\beta)},
\end{equation}
where $g(\beta)\equiv \beta^{p-1}(1-\beta)>0$.
$g(\beta)$ has the maximum value at $\beta =\frac{p-1}{p}$
and
\begin{equation}
 g(\frac{p-1}{p})=\frac{1}{p}(\frac{p-1}{p})^{p-1}.
\end{equation}
Thus,
\begin{equation}
 x_p(a_{ p,max })>
\frac{1}{p}
(\frac{p-1}{p})^{p-1}\frac{\beta}{g(\frac{p-1}{p})}
=\beta.
\end{equation}

\section{Proof of $H(x_1)H(x_2) \cdots H(x_{p-1})
= RL\cdots L$ for $a_{ p,0 }\le a$}

In this appendix, we prove that for $a_{ p,0 }\le a$,
the symbolic sequence for  $x_1, x_2, \cdots, x_{p-1}$
is  $RL\cdots L$.  We prove this in the asymmetric case.
We obtain the results for the symmetric case by putting
$\gamma=1$.

First, we prove the following relation,
\begin{equation}
a_{ p,0 }>\frac{\beta}{\alpha}.
\end{equation}
For $p \ge 3$, the following relation holds,
\begin{equation}
x_p(\frac{\beta}{\alpha})=
(\frac{\beta}{\alpha})^{p-2} \beta >\beta.
\end{equation}
Since $x_p(a)=\gamma a^{p-1} ( 1- a \alpha)$
is a strictly decreasing continuous function
for $a_{p,max} \le a$, and $a_{p,max}< a_{p,0}$, 
we obtain $a_{ p,0 }>\frac{\beta}{\alpha}$.\\
Now, let us fix $a$ such as $a\ge a_{p,0}$.  Then,
from the relation (D$\cdot$1)
\[x_1=a\alpha \ge a_{p,0} \alpha > \beta.\]
Thus, $H(x_1)=R$.  Then,
\[x_2=\gamma a ( 1- x_1)= \frac{
\gamma a^{p-1} ( 1- a \alpha)}{a^{p-2}}
=\frac{x_p}{a^{p-2}}\le \frac{\beta}{a^{p-2}} 
\le \frac{\beta}{a_{p,0}^{p-2}} <\frac{\beta}{a_{p,0}}
<\alpha.\]
Thus, $x_2 \in I_L$.
Now, let us assume $x_n<\beta$ for $2\le n \le p-2$.
Then, 
\[ x_{n+1}=ax_n =a^n \gamma(1-a\alpha)=
\frac{a^{p-1} \gamma(1-a\alpha)}{a^{p-1-n}}<
\frac{\beta}{a^{p-1-n}}<\frac{\beta}{a}\le 
\frac{\beta}{a_{p,0}}<\alpha.
\]
Therefore, $x_{n+1} \in I_L$.
Thus, 
\[x_2<x_3<\cdots<x_{p-1}<\alpha.\]
Q.E.D.\\

\section{Proof of 
$a_c \le a_{3,0}<a_{4,0}<\cdots < 
a_{p,0}< a_{p+1,0}<\cdots <a_M$}

In this appendix, we prove the following relations,
\[ a_c \le a_{3,0}<a_{4,0}<\cdots < a_{p,0}< a_{p+1,0}<\cdots
<a_M.
\]
We only have to prove it for the asymmetric case
because $\gamma =1$ reduces to the symmetric case.

Let us estimate $x_{p+1}(a_{p,0})$ for $p \ge 3$,
\[ x_{p+1}(a_{p,0})=a_{p,0}x_p(a_{p,0})=
a_{p,0}\beta>\beta.
\]
Since $x_{p+1}(a)$ is decreasing for 
$a_{p+1,max} \le a$ and $a_{p+1,max}<a_{p+1,0}$,
$a_{p,0}<a_{p+1,0}$ follows.\\
Let us prove the first relation $a_c \le a_{3,0}$.
For $a_{3,0}<a<a_{3,1}$, there is a stable periodic 
three solution.  
If $a_c>a$, there is a stable periodic solution
with period $2^m$ with some integer $m$,
and no other stable solution.
Thus, $a_c \le a_{3,0}$.
Q.E.D.


\end{document}